\documentclass{emulateapj}

\usepackage{natbib}

\usepackage{graphicx}
\usepackage{latexsym}
\usepackage{amsfonts,amsmath,amssymb}
\usepackage{url}
\usepackage[utf8]{inputenc}
\usepackage{fancyref}
\usepackage{hyperref}
\hypersetup{colorlinks=false,pdfborder={0 0 0},}

\def\msol{$M_{\odot}$}

\def\lya{Ly$\alpha$}

\shorttitle{Are Ultra-faint Galaxies at $z=6-8$ Responsible for Cosmic Reionization ?}
\shortauthors{Atek et al.}

\begin{document}


\title{Are Ultra-faint Galaxies at $z=6-8$ Responsible for Cosmic Reionization ? \\
Combined constraints from the Hubble Frontier Fields Clusters and Parallels \footnotemark[$\dagger$]}

\footnotetext[$\dagger$]{Based on observations made with the NASA/ESA Hubble Space Telescope, which is operated by the Association of Universities for Research in Astronomy, Inc., under NASA contract NAS 5-26555. These observations are associated with programs 13495, 11386, 13389, and 11689. STScI is operated by the Association of Universities for Research in Astronomy, Inc. under NASA contract NAS 5-26555. The Hubble Frontier Fields data and the lens models were obtained from the Mikulski Archive for Space Telescopes (MAST). This work utilizes gravitational lensing models produced by PIs Ebeling, Merten \& Zitrin, funded as part of the HST Frontier Fields program conducted by STScI.}

\author{Hakim Atek\altaffilmark{1,2}}
\author{Johan Richard\altaffilmark{3}}
\author{Mathilde Jauzac\altaffilmark{4,5}}
\author{Jean-Paul Kneib\altaffilmark{1,6}}
\author{Priyamvada Natarajan\altaffilmark{2}}
\author{Marceau Limousin\altaffilmark{6}}
\author{Daniel Schaerer\altaffilmark{7,8}}
\author{Eric Jullo\altaffilmark{6}}
\author{Harald Ebeling\altaffilmark{9}}
\author{Eiichi Egami\altaffilmark{10}}
\author{Benjamin Clement\altaffilmark{3}}

\altaffiltext{1}{Laboratoire d'Astrophysique, Ecole Polytechnique F\'ed\'erale de Lausanne, Observatoire de Sauverny, CH-1290 Versoix, Switzerland}
\altaffiltext{2}{Department of Astronomy, Yale University, 260 Whitney Avenue, New Haven, CT 06511, USA}
\altaffiltext{3}{CRAL, Observatoire de Lyon, Universit\'e Lyon 1, 9 Avenue Ch. Andr\'e, 69561 Saint Genis Laval Cedex, France}
\altaffiltext{4}{Institute for Computational Cosmology, Durham University, South Road, Durham DH1 3LE, U.K}
\altaffiltext{5}{Astrophysics and Cosmology Research Unit, School of Mathematical Sciences, University of KwaZulu-Natal, Durban, 4041 South Africa}
\altaffiltext{6}{Aix Marseille Universit\'e, CNRS, LAM (Laboratoire d'Astrophysique de Marseille) UMR 7326, 13388, Marseille, France}
\altaffiltext{7}{Observatoire de Gen\`eve, Universit\'e de Gen\`eve, 51 Ch. des Maillettes, 1290, Versoix, Switzerland} 
\altaffiltext{8}{CNRS, IRAP, 14 Avenue E. Belin, 31400, Toulouse, France}
\altaffiltext{9}{Institute for Astronomy, University of Hawaii, 2680 Woodlawn Drive, Honolulu, Hawaii 96822, USA}
\altaffiltext{10}{Steward Observatory, University of Arizona, 933 North Cherry Avenue, Tucson, AZ, 85721, USA}

\begin{abstract}

We use deep {\em Hubble Space Telescope} ({\em HST}) imaging of the Frontier Fields (FF) to accurately measure the galaxy rest-frame ultraviolet luminosity function (UV LF) in the redshift range $z \sim 6-8$. We combine observations in three lensing clusters A2744, MACS0416, MACS0717 and their associated parallels fields to select high-redshift dropout candidates. We use the latest lensing models to estimate the flux magnification and the effective survey volume in combination with completeness simulations performed in the source plane. We report the detection of 227 galaxy candidates at $z=6-7$ and  25 candidates at $z \sim 8$. While the total survey area is about 4 arcmin$^{2}$ in each parallel field, it drops to about 0.6 to 1 arcmin$^{2}$ in the cluster core fields because of the strong lensing. We compute the UV luminosity function at $z \sim 7$ using the combined galaxy sample and perform Monte Carlo simulations to determine the best fit Schechter parameters. We are able to reliably constrain the LF down to an absolute magnitude of $M_{UV}=-15.25$, which corresponds to 0.005$L^{\star}$. More importantly, we find that the faint-end slope remains steep down to this magnitude limit with $\alpha=-2.04_{-0.17}^{+0.13}$. We find a characteristic magnitude of $M^{\star} = -20.89_{-0.72}^{+0.60}$ and Log($\phi^{\star}$)=$-3.54_{-0.45}^{+0.48}$. Our results confirm the most recent results in deep blank fields \citep{finkelstein14, bouwens14} but extend the LF measurements more than two magnitudes deeper. The UV LF at $z \sim 8$ is not very well constrained below $M_{UV}=-18$ due to the small number statistics and incompleteness uncertainties. To assess the contribution of galaxies to cosmic reionization we derive the UV luminosity density at $z\sim7$ by integrating the UV LF down to an observational limit of $M_{UV} = -15$. We show that our determination of Log($\rho_{UV}$)=$26.2\pm0.13$ (erg s$^{-1}$ Hz$^{-1}$ Mpc$^{-3}$) can be sufficient to maintain reionization with an escape fraction of ionizing radiation of $f_{esc}=10-15$\%. Future HFF observations will certainly improve the constraints on the UV LF at the epoch of reionization, paving the way to more ambitious programs using cosmic telescopes with the next generation of large aperture telescopes such as the {\em James Webb Space Telescope} and the {\em European Extremely Large Telescope}.
 \vspace{0.3cm}
\end{abstract}


\section{Introduction} 

One of the most important challenges in observational cosmology is the identification of the sources responsible for cosmic reionization. Shortly after the Big Bang the Universe was completely neutral following the recombination of hydrogen atoms, until the first sources started to reionize the neutral gas in their surroundings. Several observational results have now narrowed down the period of reionization to the redshift interval $6 <z<12$. Observations of the Gunn-Peterson effect in the absorption spectra of quasars and gamma-ray bursts (GRB) indicate that the Universe was mostly ionized by $z \sim 6$ \citep{fan06,chornock14}. The sensitivity of \lya\ emission to neutral gas is also used to probe the ionization state of the intergalactic medium (IGM) at $z>6$. In particular, the prevalence of \lya\ emitters (LAEs) among continuum selected galaxies (LBGs) appears to drop very rapidly at $z > 6.5$, which suggests an increase in the fraction of neutral hydrogen \citep{stark10, treu13, schenker14, pentericci14}. The decline in the LAEs fraction also favors a patchy rather than a smooth reionization process. The optical depth of Thomson scattering to the cosmic microwave background (CMB) recently reported by the Planck collaboration \citep{planck15} implies a redshift of instantaneous reionization around $z_{r}=8.8_{-1.2}^{+1.3}$, significantly lower than earlier determinations of $z_{r} = 10.6 \pm 1.1$ by the Wilkinson Microwave Anisotropy Probe \citep[WMAP,][]{bennett13}. Despite these major advances, large uncertainties remain regarding the main sources that drive the reionization process.


Early star-forming galaxies are now thought to be the best candidates for providing the required ionizing power \citep[e.g.][]{robertson14,mitchell15, duncan15,finkelstein14,mason15}. Deep imaging campaigns of blank fields with the {\em Hubble Space Telescope} and ground-based instrumentation have made important inroads in constraining the galaxy ultraviolet luminosity function (UV LF) out to $z \sim 10$ \citep[e.g.,][]{bouwens12,bunker10,oesch12,mclure13,schmidt14,bouwens14,finkelstein14}, which, in turn, encodes important information about the cosmic star formation \citep{bouwens14, robertson15}. Indeed, the rest-frame UV radiation traces the recent star formation averaged over the past hundreds Myr. Therefore, the integration of the UV LF provides the luminosity density of galaxies at a given redshift, which can be translated to the ionizing radiation, assuming a certain star-formation history. The UV luminosity density is sensitive to two main parameters: (i) the faint-end slope of the LF, and (ii) the integration limit at the faint end. 

The faint-end slope of the LF evolves with redshift, and gets as steep as $\alpha \sim -2$ at $z \sim 7$. For comparison, the faint-end slope at $z < 0.5$ slope is very shallow ($\alpha \sim -1.3$), which indicates a decreasing contribution of faint star-forming galaxies to the total star formation density towards lower redshift \citep{arnouts05,schiminovich05}. The predicted evolution of the dark matter halo mass function based on cosmological simulations predicts an even steeper slope of $\alpha \sim 2.3$ of the UV LF at high redshift \citep{jaacks12}. The deepest {\em HST} observations of blank fields, such as the XDF, put constraints on the faint-end slope down to an absolute magnitude of $M_{UV} \sim -17.5$ AB. However, the ability of galaxies to reionize the Universe relies on the extent of the steep faint-end slope down to lower luminosities, typically around 0.001L$^{\star}$ at $z \sim 7$\footnote{The characteristic magnitude $M_{UV}^{\star}$=-21 AB \citep[e.g.][]{bouwens14,atek14b}}. While cosmological simulations point to a halo mass limit of $10^{6}$\msol\ for early galaxy formation \citep{jaacks12,kimm14,wise14}, the depth of current observations however limits the exploration of the LF to galaxies brighter than $\sim$0.1L$^{\star}$. One particularly efficient way to push the limits of current facilities is to take advantage of the gravitational lensing offered by massive galaxy clusters, which act as natural telescopes by boosting the flux of background sources \citep{kneib11}. 

Since the discovery of the first giant arcs created by strong lensing \citep{soucail87}, cosmic lenses have been successfully used to detect intrinsically faint background sources and perform spatially detailed analysis of distant galaxies. Recently, using HST imaging of 25 X-ray-selected clusters in the Abell \citep{abell89} and MACS \citep{ebeling01,ebeling07,ebeling10,mann12} catalogs, the CLASH program \citep{postman12} has made important progress in the characterization of the lensing properties of clusters. Multi-wavelength observations has enabled the measurement of the total cluster mass with a precision of 10\% and the detection of some of the most distant galaxies at $z>7$ \citep[e.g.][]{zitrin11, bradley14, coe13}. However, two limitations prevented such program from exploring the faint-end of the LF at the epoch of reionization. First, the relatively shallow data compared to HUDF, restricted the survey to the brighter end part of the LF, even in the case of high magnifications. In this sense, wide area surveys, such as BoRG \citep{trenti11}, are indeed well suited to help constrain the brighter part of the LF \citep{bradley12,schmidt14}. Second, the lack of very accurate lensing models for some of these clusters, mostly due to the low number of multiple images available to constrain the mass distribution, also thwarts the construction of a reliable LF. 

The Hubble Frontier Fields (HFF) project aims at overcoming these two major limitations by obtaining deep multi-wavelength observations of six massive galaxy clusters that act as cosmic lenses. The HFF include the deepest optical and near-infrared observations of lensing clusters using {\em HST} director discretionary time, which are complemented by a wealth of data including {\em ALMA, Spitzer, Chandra, XMM, VLA}, as well as {\em HST} and ground-based imaging and spectroscopic follow-up \citep[e.g.][]{owers11,ebeling14,rawle15,karman15,richard14b,medezinski15,rodney15,zitrin15,schirmer15,grillo15,treu15,ogrean15}. Based on the full HFF dataset of the first cluster A2744, \citet{atek15} presented the first constraints on the UV LF at $z\sim 7$ and $z \sim 8$ \citep[see also][]{ishigaki15}. The key result was the steep faint-end slope of $\alpha \sim -2.01$ that extends down to M$_{UV} = -15.5$ AB. However, large uncertainties arising from small sample size, lensing models, and cosmic variance still prevent strong conclusions on the total UV luminosity density of galaxies at the epoch of reionization. 

In this paper, we combine the complete dataset of the three lensing clusters A2744, MACS0416, and MACS0717, and their respective parallel fields to search for high-redshift dropout galaxies and put stronger constraints on the UV LF at $z > 6$. The paper is organized as follows. In Section \ref{sec:obs}, we describe the observations. The sample selection method is described in Section \ref{sec:sample}. In Section \ref{sec:models}, we briefly describe the lensing models and the multiple-image identification. The procedure and the results of the computation of the UV luminosity function are presented in Section \ref{sec:lf}. A summary is given in Section \ref{sec:summary}. Throughout the paper, we adopt a standard $\Lambda$CDM cosmology with $H_0=71$\ km s$^{-1}$\ Mpc$^{-1}$, $\Omega_{\Lambda}=0.73,$\ and $\Omega_{m}=0.27$ to be consistent with previous studies. All magnitudes are expressed in the AB system.

\vspace{1cm}

\section{HFF Observations}
\label{sec:obs}
The HFF clusters and parallel fields were observed by {\em HST} with three ACS (Advanced Camera for Survey) optical filters (F435W, F606W, F814W) and four WFC3 (Wide Field Camera 3) near-IR filters (F105W, F125W, F140W, F160W). Observations were scheduled in two-epoch sequences, obtaining ACS observations of the main cluster and WFC3 observations of the parallel field in one epoch and swapping instruments in the second epoch. We use the high-level science products delivered by the Space Telescope Science Institute (STScI) through the Mikulski Archive for Space Telescopes\footnote{\url{https://archive.stsci.edu/prepds/frontier/}} (MAST), which include drizzled science and weight images. 

Basic calibrations were performed with the standard IRAF procedures CALACS and CALWF3 for ACS and WFC3 data, respectively. Here we chose a pixel scale of 60 mas~pix$^{-1}$ for both optical and IR drizzled images. For the ACS bands we used the "self calibrated" mosaics, which contain additional corrections applied by the STScI team to better account for charge transfer inefficiency (CTI) effects. Similarly, the WFC3 bands were also corrected for a time-variable IR background. We refer the reader to a detailed explanation of the data reduction performed by STScI\footnote{\url{http://www.stsci.edu/hst/campaigns/frontier-fields/}} (Koekemoer et al. in prep). For the fields that have been observed prior to the HFF program, we combine all the available data using the weight maps included in the HFF data release. Table \ref{tab:obs} summarizes the exposure times and the depth achieved in each filter for each of the fields. The limiting magnitude in each fitter was calculated using 0.4\arcsec diameter apertures randomly distributed in the image to sample the sky variance before fitting the resulting distribution. The quoted depth is given at the 3-$\sigma$ level.

\begin{table*}
\caption{{\em HST} observations of the HFF fields}
\label{tab:obs} 
\begin{tabular}{lccccccccc}
\hline
 & & A2744  &  & & MACS0416 & & & MACS0717 & \\ \hline
 Filter & \#Orbits & Depth\footnote{The depth of the images are 3-$\sigma$ magnitude limits measured in a 0.4\arcsec aperture.} & Obs Date  & \#Orbits & Depth & Obs Date & \#Orbits & Depth & Obs Date \\ \hline 
 WFC3/F160W  & 24  & 28.3&Oct/Nov 2013       & 24  & 29.1&Jul/Sep  2014   & 26  & 28.8&Feb/Mar  2015\\
 WFC3/F140W  & 10  & 29.1&Oct/Nov 2013       & 10  & 28.8&Jul/Sep  2014   & 12  & 28.5&Feb/Mar  2015 \\
 WFC3/F125W  & 12  & 28.6&Oct/Nov 2013       & 12  & 28.8&Jul/Sep  2014   & 13  & 28.6&Feb/Mar  2015 \\
 WFC3/F105W  & 24  & 28.6&Oct/Nov 2013       & 24  & 29.2&Jul/Sep  2014   & 27  & 28.9&Feb/Mar  2015 \\
 ACS/F814W    & 42  & 29.4 &Jun/Jul 2014        & 50  & 29.2&Jan/Feb 2014   & 46  & 29.3&Sep/Nov 2014\\ 
 ACS/F606W    & 10  & 29.4&Jun/Jul 2014         & 13  & 29.1&Jan/Feb 2014   & 11  & 28.6&Sep/Nov 2014\\ 
 ACS/F435W    & 18  & 28.8 &Jun/Jul 2014        & 21  & 30.1&Jan/Feb 2014   & 19  & 29.5&Sep/Nov 2014\\ \hline
 & & A2744  &  & & MACS0416  & & & MACS0717& \\ 
 & & Parallel  &  & & Parallel  & & & Parallel  \\ \hline
 Filter & \#Orbits & Depth & Obs Date  & \#Orbits & Depth & Obs Date & \#Orbits & Depth & Obs Date \\ \hline 
  WFC3/F160W  & 24  & 28.5&Jun/Jul 2014         & 28  & 28.5&Jan/Feb  2014   & 26  & 28.7&Sep/Nov 2014\\
 WFC3/F140W   & 10  & 28.2&Jun/Jul 2014         & 12  & 28.6&Jan/Feb  2014   & 10  & 28.3&Sep/Nov 2014 \\
 WFC3/F125W   & 12  & 28.34&Jun/Jlu 2014        & 12  & 28.6&Jan/Feb  2014   & 14  & 28.5&Sep/Nov 2014 \\
 WFC3/F105W   & 24  & 28.6&Jun/Jul 2014         & 28  & 28.9&Jan/Feb  2014   & 24  & 28.6&Sep/Nov 2014 \\
 ACS/F814W     & 42  & 29.0 &Oct/Nov 2013       & 42  & 28.9&Jul/Sep  2014   & 43  & 28.7&Feb/Mar  2015\\ 
 ACS/F606W     & 16  & 28.9 &Oct/Nov 2013       & 10  & 28.7&Jul/Sep  2014   & 16  & 28.4&Feb/Mar  2015\\ 
 ACS/F435W     & 28  & 29.7 &Oct/Nov 2013       & 18  & 29.1&Jul/Sep  2014   & 26  & 29.1&Feb/Mar  2015 \\  \hline
\end{tabular}
\end{table*}

\begin{figure}[!htbp]
   \centering
   \includegraphics[width=8.5cm]{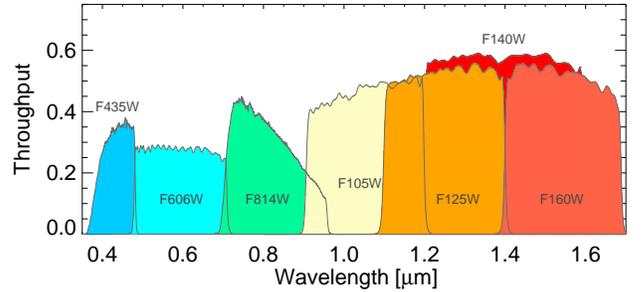} 
   \caption{Throughput curves of the HST/ACS and WFC3 filter set used for the HFF observations.}
   \label{fig:example}
\end{figure}

\section{Sample Selection}
\label{sec:sample}

\subsection{Photometric catalogs}

We constructed the photometric catalogs in each field using the {\tt SExtractor} software \citep{bertin96}. We first matched all the images to the same point spread function (PSF) using a model based on the largest PSF derived with {\tt TinyTim} \citep{Krist_Hook_Stoehr_2011}. In order to increase the sensitivity to faint sources, we created deep images by using inverse variance (IVM) weight maps to combine all IR frames for $z \sim 7$ sources and F125W, F140W and F160W for $z \sim 8$ sources, respectively. This deep image is used for source detection in the SExtractor dual image mode, while individual images are used for photometry, weighted by the individual IVM images. 

One important limitation for the detection of the faintest sources in the cluster fields is the contamination from the intra cluster light (ICL) in the central region of the cluster. This diffuse light, concentrated primarily in the central region of the cluster, is due to the brightest cluster galaxy (BCG) or other bright cluster members and the tidal stripping of stars from interacting galaxies during the merging history of the cluster \citep[e.g.][]{montes14}. In order to mitigate the contamination of sources close to the cluster center we subtract a median filtered image from the detection one with a filter size of $\sim $ 2\arcsec$\times 2$\arcsec , while the photometry is performed on the original image. The adopted filter size is a tradeoff between the removal of extended bright emission and the appearance of artifacts close the bright galaxies of the cluster core. {We note that the median filtered detection image allows us to detect five more sources on average in cluster fields compared to the original images. This is also important for the visual inspection of galaxy candidates as the background flux is much lower in these corrected images.} 

In order to estimate the total flux errors, we ran simulations of galaxies with different sizes and profiles (cf. the completeness simulations in Section \ref{sec:comp}) and compared the recovered fluxes to the input values. We find that the median-filtering approach achieves an uncertainty of 0.5 mag for the faintest sources ($H_{140} \sim 28-29$ mag), by using a local background estimate with {\tt back\_size = 6} in {\tt SExtractor} for photometry. In addition, the following extraction parameters were set to improve the detection of the faintest sources in the field: {\tt detect\_minarea}=2 and {\tt detect\_thresh=1.5} for the detection and {\tt deblend\_nthresh=16} for source deblending. Two types of photometry are used throughout this work. The isophotal magnitude (ISO) is used to compute the colors and ensure that the same aperture is used across the filter set. The total flux is measured within the Kron radius using the AUTO magnitude, which hereafter is used as the total magnitude. We modified the magnitude errors to account for pixel-to-pixel noise correlations in the drizzled images following \citet{r_Hook_Levay_Lucas_et_al__2000}. Finally, the individual catalogs were matched into a master photometric catalog and cleaned from spurious sources.

\begin{figure*}[!htbp]
   \centering
   \includegraphics[width=6cm]{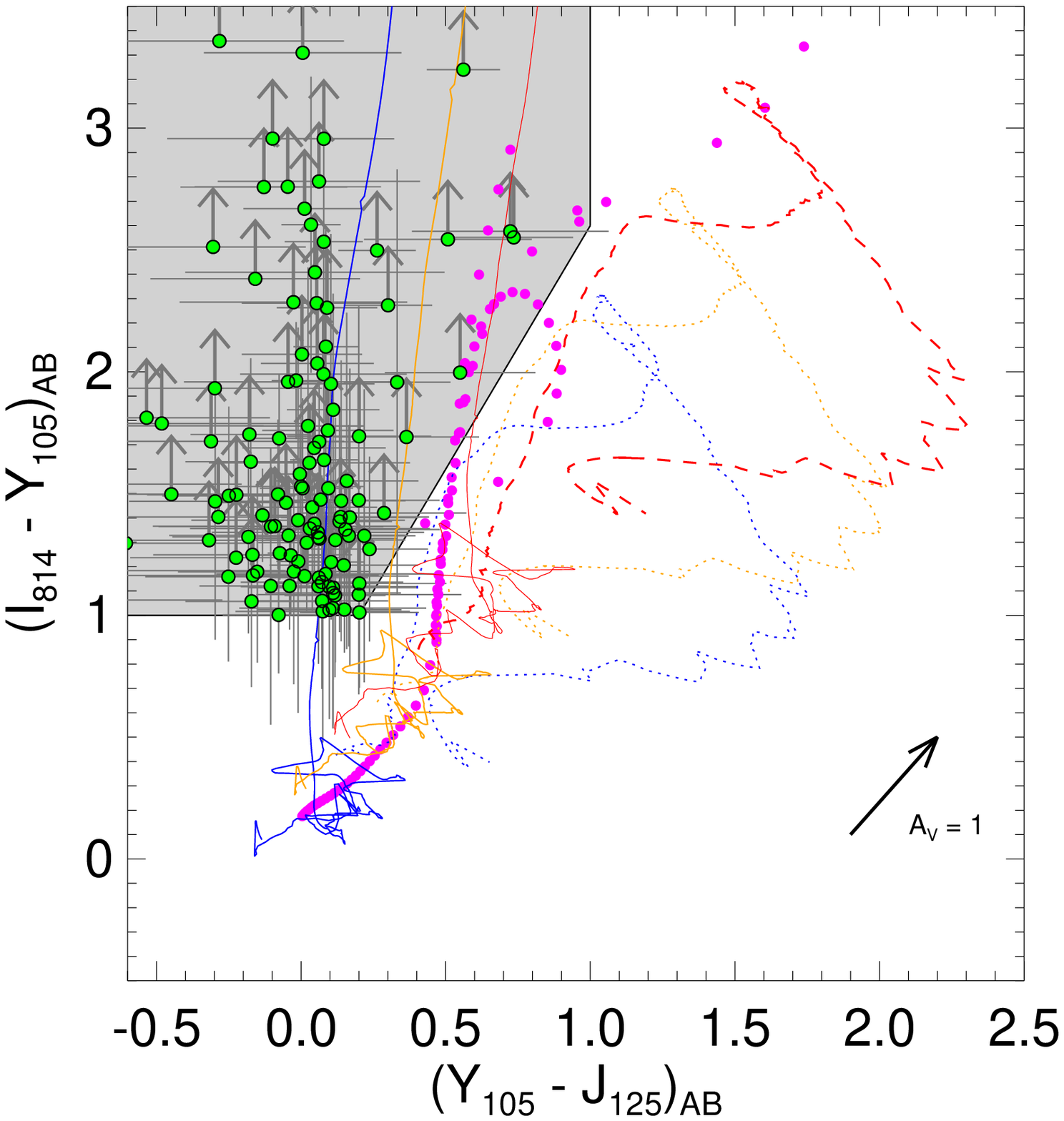} 
   \includegraphics[width=6cm]{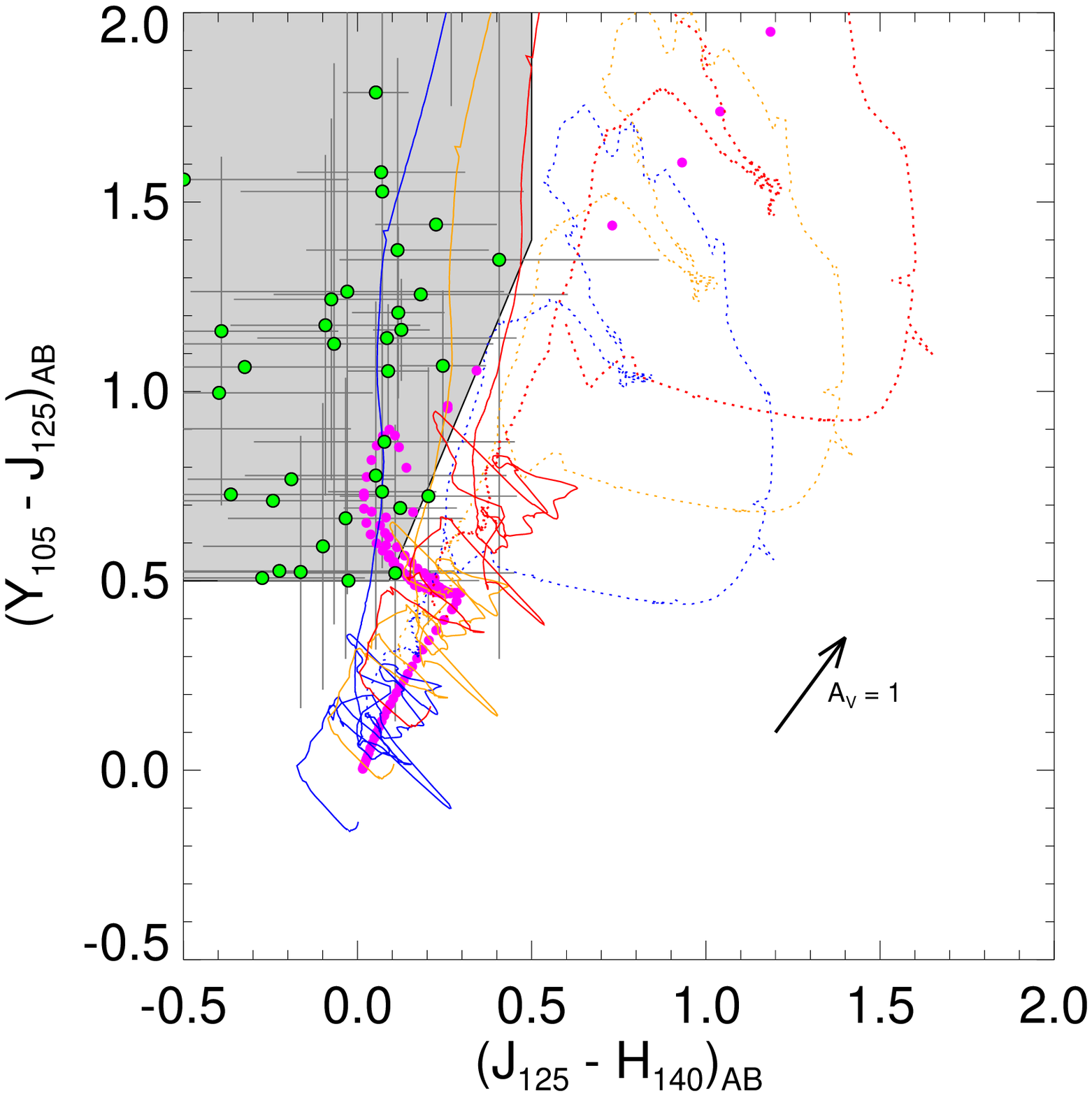} 
    \includegraphics[width=6cm]{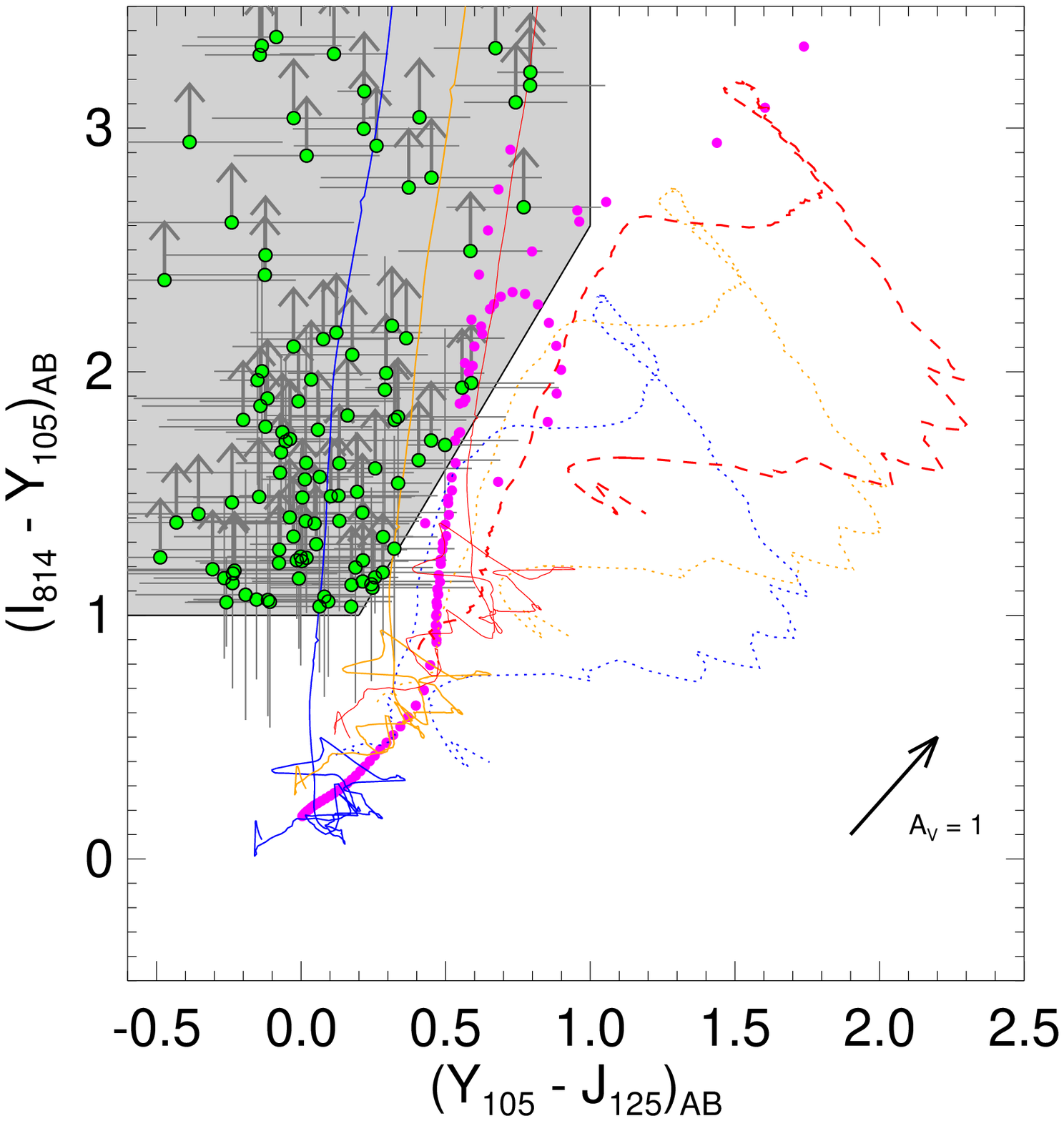} 
     \includegraphics[width=6cm]{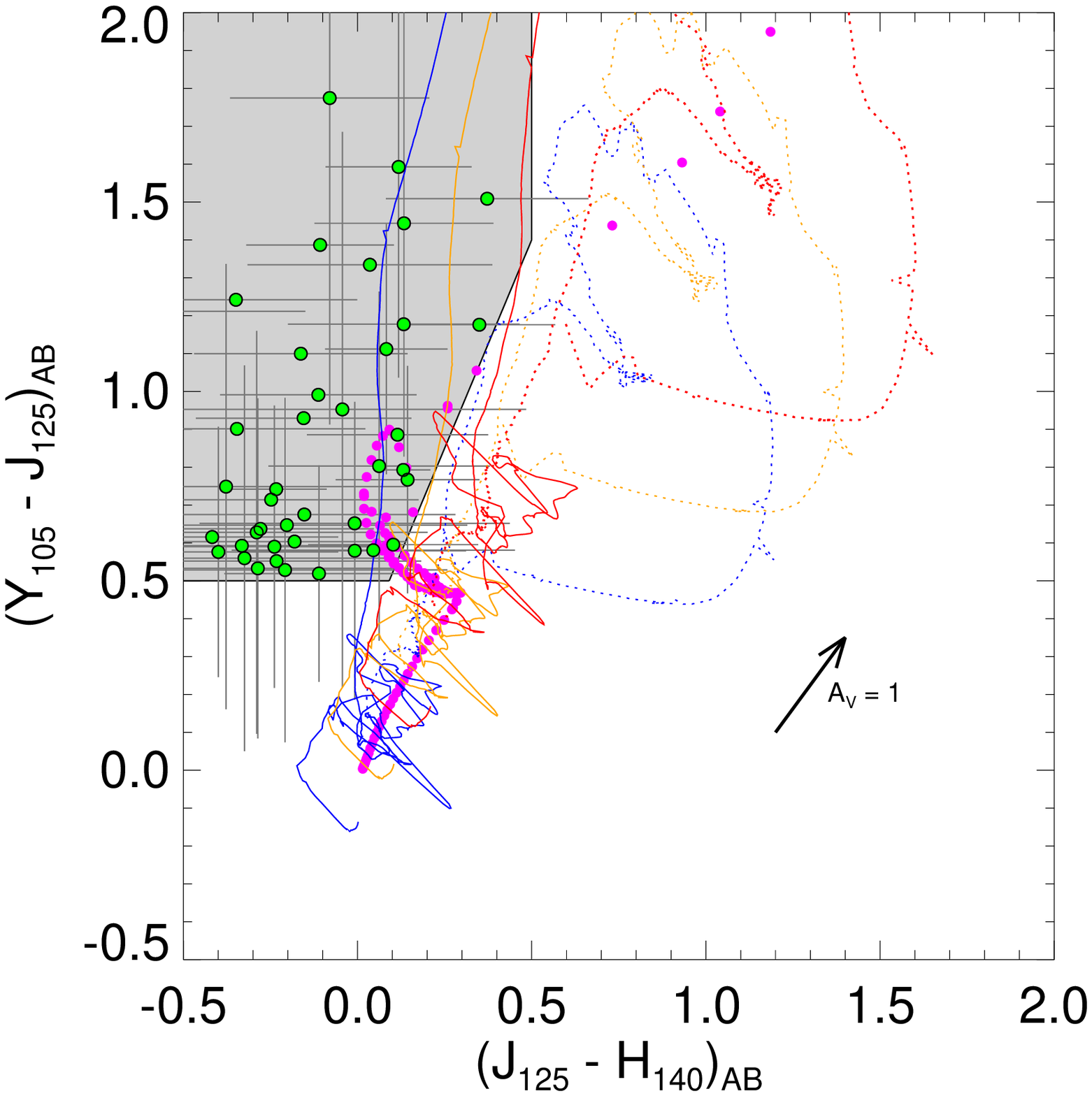} 
   
   \caption{The Color-color selection windows (represented by the shaded regions) of high-$z$ dropout candidates. The left and right panels show the selection of $z \sim 7$ and $z \sim 8$ galaxies, respectively. The top panels are for cluster fields and bottom panels for parallel fields. The green circles (with associated 1-$\sigma$ uncertainties) represent a compilation of all galaxies satisfying our selection criteria and included in our $z > 6$ samples. The dotted line show the color track, i.e. the redshift-evolution of colors, of low-redshift elliptical galaxies generated from \citet{coleman80} templates, whereas solid lines show starburst galaxies generated from \citet{kinney96} templates. The color code, from blue to red, illustrates the impact of extinction in steps of $A_{V}=1$. Finally, the magenta points denotes the color track of stars generated from \citet{chabrier00} templates.}
   \label{fig:color}
\end{figure*}

\subsection{High-redshift Dropout Selection}

We adopted the Lyman break selection technique \citep{steidel96,giavalisco02} to detect high-redshift galaxy candidates. The selection is based on color-color criteria to sample the UV continuum dropout of the IGM absorption blueward of \lya\ caused by the intervening hydrogen along the line of sight, and minimize contamination by low-redshift red objects at the same time. Following \citep{atek15}, we use the following criteria to select $z=6-7$ galaxies:

\begin{align}
\label{eq:criteria7}
 (I_{814} {-} Y_{105})   &>  1.0  \notag \\
(I_{814} {-} Y_{105})   &>  0.6 + 2.0(Y_{105} {-} J_{125})\\
(Y_{105} {-} J_{125})  &< 0.8 \notag
\end{align}

In addition, we require all sources to be detected in the deep IR image and in at least two IR bands with 4-$\sigma$ significance or higher. We also reject any galaxy that shows up at a significant (at $1.5 \sigma$) level in the deep optical combination of $B_{435} + V_{606}$ images. To preserve the Lyman break criterion in the case of non detection in the $I_{814}$ image, we assign a 2-$\sigma$ limiting magnitude to this band. Therefore, we select only galaxies with $Y_{105}$ at least 1 mag brighter than the $I_{814}$ depth. Similarly, we select $z \sim 8$ galaxies that satisfy:

\begin{align}
\label{eq:criteria8}
(Y_{105} {-} J_{125})  &>  0.5 \notag \\
(Y_{105} {-} J_{125})  &>  0.3 + 1.6(J_{125} {-} H_{140})\\
(J_{125} {-} H_{140})  &< 0.5 \notag
\end{align}

We require a 4-$\sigma$ detection in a deep $J_{125}+H_{140}+H_{160}$ image and the detection in the stacked deep optical image (four ACS filters) to be less than 1.5$\sigma$. The deep optical image is about one magnitude deeper than the faintest sources (detected in the IR) accepted in our sample. In Figure \ref{fig:color} we show the result of our selection procedure. The green circles show the location of all high-redshift galaxies identified in this work in the color-color diagram. The dotted and solid lines represent the color evolution of low-redshift elliptical galaxies and high-redshift starbursts, respectively, constructed from \citet{coleman80} and \citet{kinney96} templates. An attenuation of $A_{V}=1, 2, 3$ is applied to the blue, orange, and red curves, respectively. The shaded region represent our adopted selection window, which was chosen to minimize contamination from low-redshift interlopers and red objects such as cool stars represented by magenta points. Our final sample, combining cluster and parallel fields, contains 227 galaxies at $z \sim 6-7$, according to the selection based on Eq. \ref{eq:criteria7}, and 25 galaxies at $z \sim 8$ based on the selection of Eq. \ref{eq:criteria8}. We find roughly the same number of candidates in the cluster and the parallel fields. While the survey area is smaller in the lensed field, the magnification bias balances the number density, allowing the detection of much fainter galaxies than in the parallel fields.

\begin{table}
\centering
\caption{Number of galaxy candidates in each field}
\label{tab:obs} 
\begin{tabular}{lcc}
\hline
Field & $z=6-7$ & $z = 8$ \\ \hline
A2744 & 45 & 7 \\
MACS0416 & 33 & 3 \\
MACS0717 & 41 & 3 \\
A2744 par & 44 & 3 \\
MACS0416 par & 33 & 5 \\
MACS0717 par & 31 & 5 \\ \\ \hline
\end{tabular}
\end{table}

\begin{figure}[!htbp]
   \centering
   \includegraphics[width=7cm]{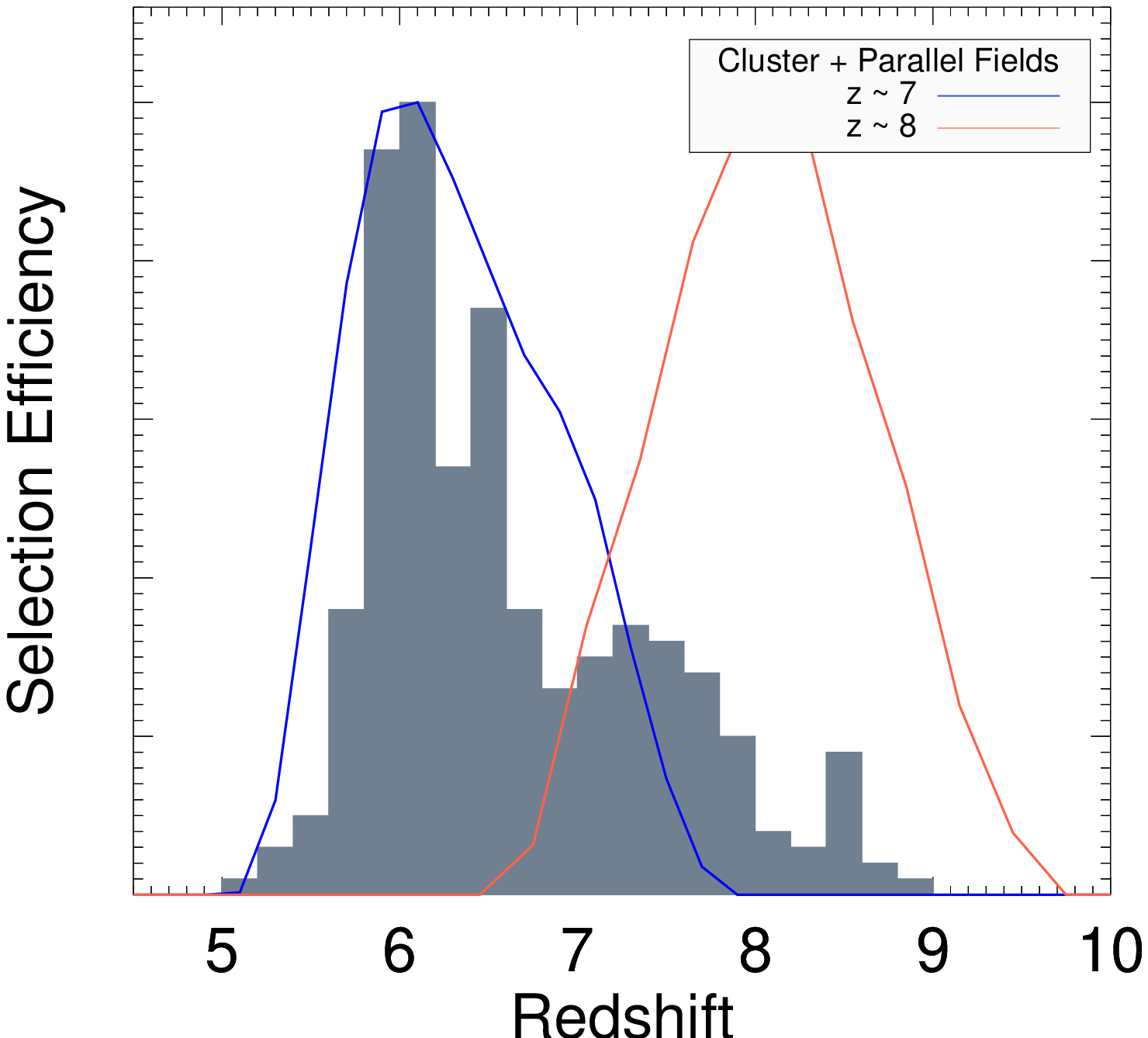} 
   \caption{The redshift selection function for all the fields at $z \sim 7$ and $z \sim 8$. The curves are the result of our completeness simulations in recovering the input sources marginalized over the redshift, while the shaded histograms represent the actual selected galaxies in our sample. The median redshift of the our $z \sim 8$ is significantly lower than expected due to the depth of the observations and the redshift evolution of the luminosity function.}
   \label{fig:example}
\end{figure}

\subsection{Sources of Contamination}
\label{sec:contamination}

We now discuss the main sources of contamination for our high-redshift sample of galaxies. A first possibility is spurious sources produced by detector artifacts, diffraction features, or photometric errors scattering into the color selection space. We have visually inspected all of our candidates to identify such contaminants. Most of the spurious sources were already cleaned using the weight maps that exclude the frame edges and the IR blobs identified by STScI calibrations. We find that the remaining artifacts are mostly diffraction spikes of bright stars and few very bright cluster galaxies in the field. Fake sources due to photometric noise are also minimal since we require the detection in at least two different bands. 

The second potential source of contamination are low-redshift galaxies that show similar colors to high-redshift candidates. According to our stringent spectral break criteria, such sources need to show an extremely red continuum or a large Blamer/4000\AA\ break. Dust obscured galaxies would be excluded by our selection because they would exhibit relatively red colors redward of the break. The remaining galaxies that can enter the selection need to have a large Balmer break and relatively blue continuum at longer wavelengths. The existence of these peculiar objects have been discussed in \citet{hayes12} where an SED of a young burst superimposed on an old stellar population with an extreme 4000\AA\ break can mimic the Lyman break colors. Such objects should be very rare and represent only a minor contamination. Moreover, in the case of a significant contamination by these interlopers, we expect to detect the blue continuum by stacking the optical images of our candidates, which is not the case.

Alternatively, the Lyman break colors can also be mimicked by extremely strong emission lines in $z=1-3$ star-forming galaxies \citep[e.g.,][]{atek11,atek14c, vanderwel11}. In this class of galaxies the contribution of the emission line flux to the total broadband flux is about 25\% on average and can reach 85\% \citep{atek11} and can have important implications, not only for the high-redshift galaxy selection, but also on the age and stellar mass derived from SED fitting \citep{schaerer09,wilkins13, schenker14,penin15,huang15}. In our case, the contamination of one band can lead to an artificial Lyman break, because the faint continuum remains undetected blueward of the break. \citet{atek11} estimate that the optical data should be about 1 magnitude deeper than the detection band to be able to rule out such interlopers. Our criteria impose a minimum break of 0.8 mag between the object flux and the limiting magnitude of the optical band. Moreover, such a contamination should be even smaller because our stacked candidate images, which are at least 1.3 mag deeper than the candidate's flux, show no significant detection in the optical bands shortward of the break. 

In general, observationally, estimating the contamination rate of such sources proves very difficult at $z > 6$ because it hinges on large spectroscopic follow-up programs that aim at detecting the redshifted \lya\ emission line in these galaxies. In addition to the large amount of telescope time needed to reach the required depth, the absence of \lya\ emission does not exclude the high-redshift nature of the associated objects. Indeed, the increasing neutral hydrogen fraction at $z > 6$ easily absorbs and diffuses \lya\ photons\citep{atek09b,stark10,caruana14,schenker14b}. Therefore, one must rely on simulated colors based on galaxy spectral templates to estimate the contamination rate of low redshift galaxies, which thus has been found to be less than 10 \% in most studies \citep[e.g.][]{oesch10a,bouwens14}.

Another source of contamination to consider are low-mass stars. As we can see in Fig. \ref{fig:color} a significant fraction of these stars (shown in magenta) can have similar colors to high redshift galaxies. However, we have excluded any source that has a SExtractor stellarity parameter greater than 0.8 to minimize point-like objects in our sample. Also, the number density of cool stars ranges from 0.02 to 0.05 arcmin$^{2}$ derived from observations with similar or better depths \citep{bouwens14}. This translates to about 1 contaminant in the entire $z \sim 7$ and $z \sim 8$ galaxy samples. We also consider transient events as possible contaminants. Because observations are taken at two different epochs (cf. Table \ref{tab:obs}), a supernova explosion may appear in IR images and not in the optical ones and could be selected as a dropout candidates. Similar colors can also be obtained in the case where the IR data were taken first and the supernova faded until the optical images. Again, different reasons point toward a negligible contamination level from transient sources. Such objects would show point-like profiles that would be excluded by the stellarity criterion explained above and our additional visual check for unresolved sources. The detection rate of supernovae is also very small and, so far, none of SN detections in the dedicated search program of the HFF \citep[e.g.,][]{rodney15} were inadvertently selected as high-z galaxies in the different HFF studies \citep{atek14b,atek15,finkelstein14,ishigaki15}.

\section{HFF Cluster Mass Models}
\label{sec:models}

In order to exploit the full potential of the HFF cluster lenses, we first need a robust model describing the total mass distribution and the lensing properties. In an effort to provide the community with all the required lensing maps to interpret background source observations, several groups have submitted their models prior to the HFF \citep[e.g.,][]{bradac05,merten11,johnson14,richard14,grillo15,coe15}. With the availability of new deep {\em HST}, {\em Spitzer}, and spectroscopic observations, the models were significantly improved thanks in particular to the discovery of a large number of multiple images \citep{diego14, grillo15, zitrin14, wang15, jauzac15, jauzac14b,ishigaki15}. Most notably, using a set of $\sim 180$ and $\sim 200$ multiple images in A2744 and MACS0416, respectively, \citet{jauzac15,jauzac14b} reconstructed the projected cluster mass down to a precision of $\sim 1$\%, which represents a significant improvement over pre-HFF mass models. 

In the present study, we use the most recent cluster models based on full-depth HFF observations and constructed by the CATS (Clusters As Telescopes) team. The models developed for the first two HFF clusters, namely A2744 and MACS0416, are already available on the MAST archive. However, STScI has started a new mass mapping initiative so all the teams provide HFF mass models of both clusters taking advantage of the full depth of the HFF data, but using the same inputs. These `unified' models will be submitted to STScI as part of our answer to the HFF call for the community to provide updated lensing maps\footnote{\url{https://archive.stsci.edu/prepds/frontier/lensmodels/}}. The mass modeling of the third cluster MACS0717 will be presented in details in Limousin et al. (2015).

The mass reconstruction of each of the clusters is explained in the publications listed above. Here we briefly describe the key points of the mass modeling procedure.

The CATS HFF mass models are built using the Lenstool\footnote{\url{http://projets.lam.fr/projects/lenstool/wiki}} software \citep{kneib93,jullo07,jullo09}. For the strong-lensing analysis, we are using a parametric approach that consists of modeling the cluster mass distribution using both cluster-scale and galaxy-scale halos. Galaxy-scale halos are important in the mass modeling, as multiple image configurations are impacted by their location. Indeed if a multiple image is close to a cluster galaxy, then the distortion created by the lensing effect will be a combination of both the cluster potential itself, as well as a galaxy-galaxy lensing effect due to the potential of the cluster member.The potentials describing these components are modeled using pseudo-isothermal elliptical mass distribution \citep[PIEMD,][]{eliasdottir07}.

For MACSJ0416, we used 149 of the most secure multiple images (over the 194 identified) to constrain the mass model. Our best fit mass model of the cluster inner core consists of two cluster-scale halos, well-aligned with the light peaks from the two brightest cluster galaxies (BCGs), plus 98 galaxy-scale halos, corresponding to the brightest cluster galaxies. 
For Abell 2744, we used 154 most secure multiple images (over the 181 identified). Our best-fit mass model consists of two cluster-scale halos to describe the dark matter distribution on large-scale, combined with 733 galaxy-scale halos, describing the distribution of cluster galaxies.
Both models delivered really good errors in the predictions of multiple image positions, an $rms$ of 0.68$\arcsec$ and 0.79$\arcsec$ for MACSJ0416 and Abell 2744 respectively.
For MACS J0717, we used the 140 most secure multiple images (over 163 identified) to constrain a mass model composed of 4 large scale DM haloes, plus 92 galaxy scales haloes, corresponding to the brightest cluster galaxies. More details of the SL analysis will be given in Limousin et al. (2015)

\begin{figure}[!htbp]
   \centering
   \includegraphics[width=8cm]{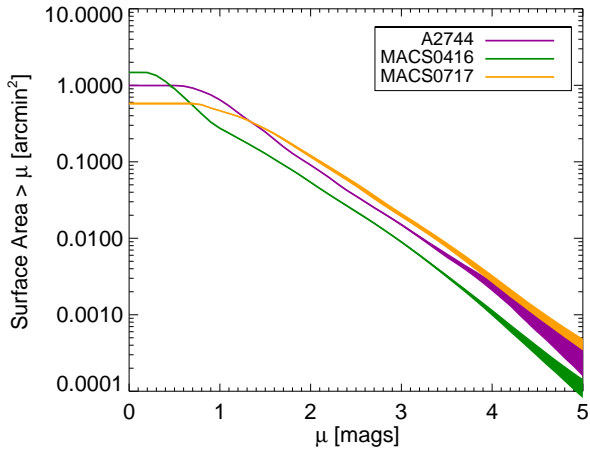} 
   \caption{The cumulative surface area in the source plane at $z \sim 7$ as a function of the amplification factor (in magnitudes) derived from the mass modeling of the three HFF clusters. Uncertainties in the surface area are also shown at the 1-$\sigma$ level .}
   \label{fig:area}
\end{figure}

\subsection{Multiple Images}

In addition to flux magnification, strong lensing produces multiple images of the same background galaxy. Therefore, we need to identify multiple images to be removed from the galaxy number counts before computing the UV LF. For each galaxy, we predict the position of potential counter images by using the mass model and Lenstool to project its position into the source plane before lensing back the source into the image plane, where Lenstool predicts the position of all the multiple images. In the vicinity of these positions, we look for dropout sources that show similar colors and photometric redshifts. In the case of well resolved sources, we also verify that they have similar morphological and geometrical symmetries constrained by the lensing model. In A2744 we find three systems at $z \sim 7$, with a total of nine multiple images, which were discussed in \citet{atek14b}. In MACS0416, we identify seven systems at $z \sim 7$. 

In the case of MACS0717, the multiple image region extends beyond the WFC3 field of view. Therefore, we expect each of the galaxy candidates to have counter-images with similar flux/magnification ratios given the lensing geometry of this cluster. The uncertainty on the image position predicted from the lensing model of MACS0717 is about $rms =1.9$ arcsec, much larger than in the two other clusters (cf. Limousin et al. 2015). Moreover, the shape of the critical line is not as well constrained, which makes it harder to identify multiple-image systems. From the Lenstool model, we estimate the multiplicity in MACS0717 to be three on average in the WFC3 field of view. Therefore, we divide the number of galaxy candidate by this multiplicity to obtain the final number counts used for the LF calculation.

\section{The Galaxy UV Luminosity Function}
\label{sec:lf}
We now turn to computing the galaxy UV luminosity function at $z \sim 7$ and $z \sim 8$ by combining the high-$z$ candidate samples from all the fields. For each field we calculate the completeness function and the effective survey volume to derive the LF. While in the parallel fields we proceed with standard completeness simulations widely used in analyzing blank field surveys, we need to take into account the lensing effects in the cluster fields.

The primarily goal of the HFF program is to extend the current limits of deep galaxy surveys by using strong lensing to magnify intrinsically faint sources behind the galaxy clusters. In parallel, the survey area in the source plane is significantly reduced for high magnifications. Consequently, the efficiency of a given lensing cluster in probing the high-redshift Universe, which can be quantified by the number of magnified sources discovered, is the result of a trade off between the amplification factor $\mu$ and the source plane area $\sigma$. We show in Fig. \ref{fig:area} the cumulative survey area as a function of the amplification factor. 

\begin{figure}[htbp]
   \centering
   \includegraphics[width=8cm]{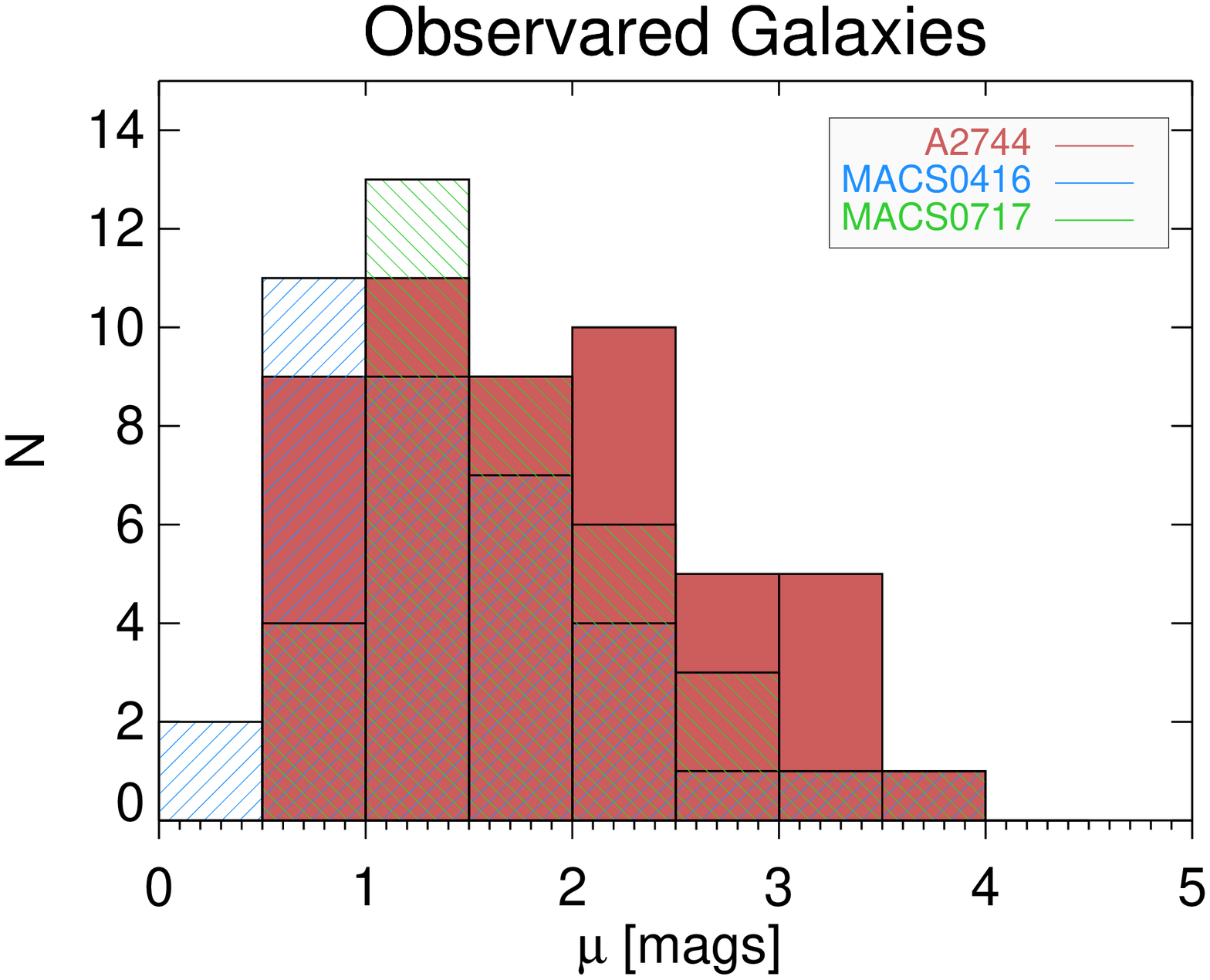} \\
 \includegraphics[width=8cm]{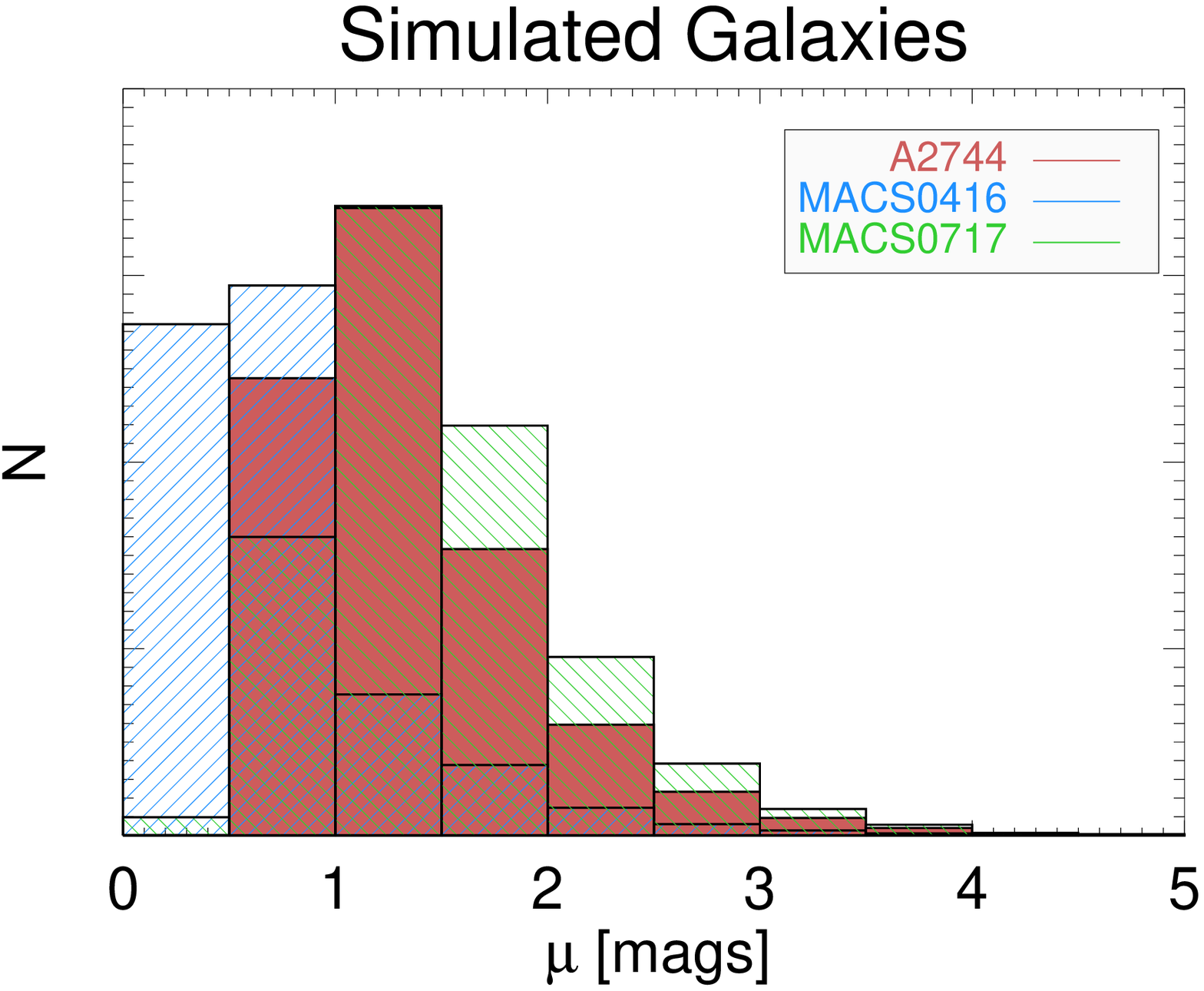}
   \caption{The distribution of the amplification factor, expressed in magnitudes. {\it Top panel} shows the distribution for the $z \sim 7$ candidates in each of the lensing clusters. For comparison,  the {\it bottom panel} presents the result of the completeness simulations in the three cluster cores, which contain about 80,000 objects each.}
   \label{fig:amp}
\end{figure}

While the survey area in the image plane corresponds to the WFC3 field of view, i.e. about 4.7 arcmin$^{2}$, we can see the total survey area is reduced to about 0.6 to 1 arcmin$^{2}$ in cluster fields. \citet{wong12} define the cluster magnification power as the cross section for magnifying a source above minimum threshold of $\mu = 3$ \citep[see also][]{richard14}. Although the precise value of the threshold is somewhat arbitrary, the number of high-redshift detections in our galaxy samples peaks around $\mu = 3$, or $\mu \sim 1.2$ mag (cf. the top panel of Fig. \ref{fig:amp}. For this typical value, the total survey area drops to values of around 0.2 to 0.5 arcmin$^{2}$ for the three clusters of this study.

The parallel fields are located about 6 arcmin away from the cluster lenses with a small yet non negligible magnification. It is important to assess the lensing effects on the UV LF, both on the magnitude and on the survey volume for these regions as well. Among the different lensing models available in the HFF project, only \citet{merten11} provide a wide-field magnification map for the three clusters that covers their flanked fields. We used the magnification maps at $z \sim 9$ that have a resolution of 25 arcsec pix$^{-1}$ to estimate the flux amplification of galaxies. Since no deflection map is available, the volume reduction is estimated by dividing the area by the amplification map. The harmonic mean of the amplification factor ranges from 1.11 to 1.23 in the three fields, with typical errors of about 10\%. The inclusion of the lensing effects introduce only small changes in the UV LF, within the error bars, basically shifting $M_{UV}$ to slightly fainter values and $\phi$ to higher values.

We now describe how we estimate the effective survey volume for each cluster by combining the source plane area with the redshift selection function and the recovery rate of simulated galaxies as a function of different galaxy and lensing parameters.

\subsection{Completeness Simulations}
\label{sec:comp}

\begin{figure}[!htbp]
   \centering
   \includegraphics[width=4cm]{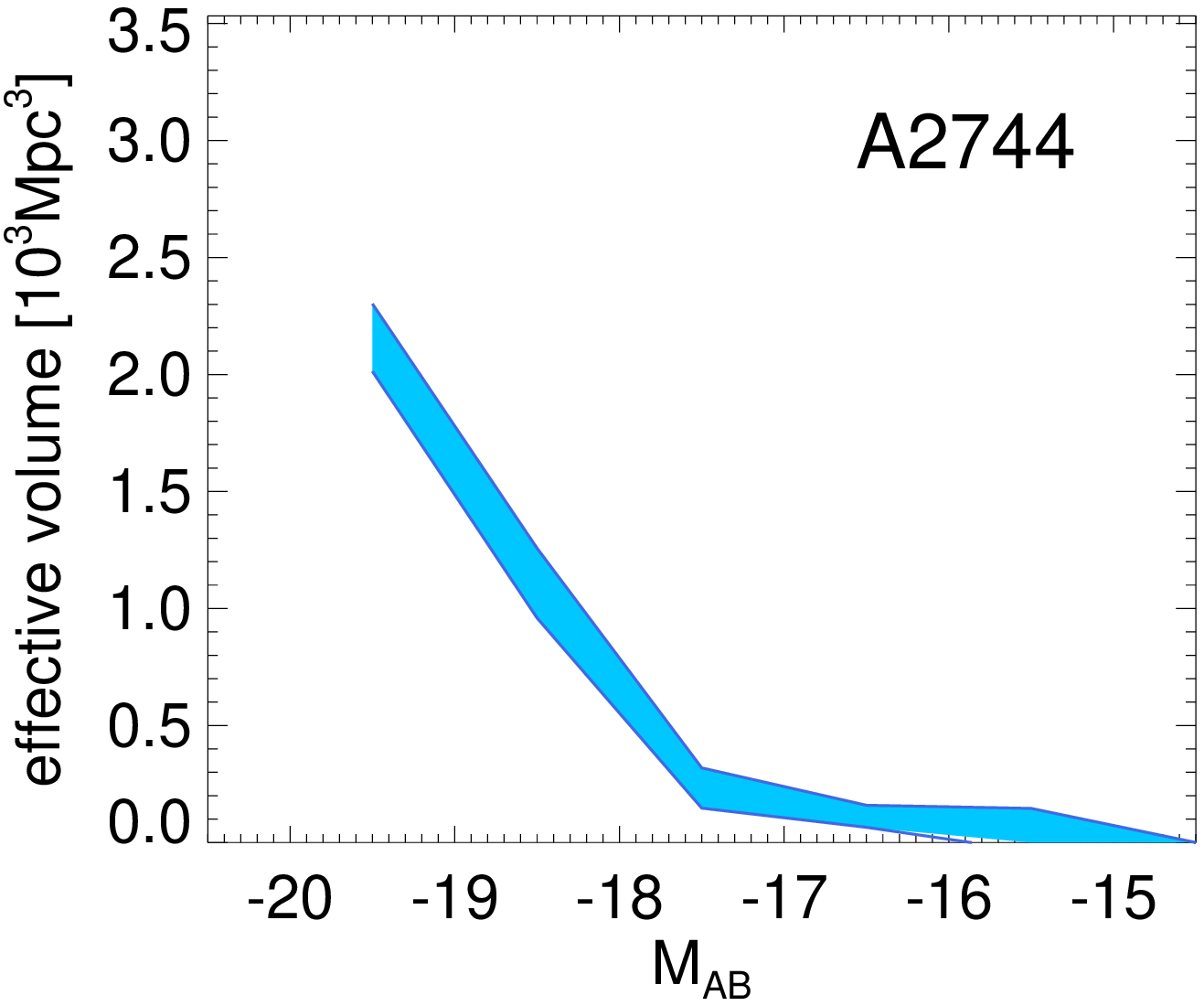}
   \includegraphics[width=4cm]{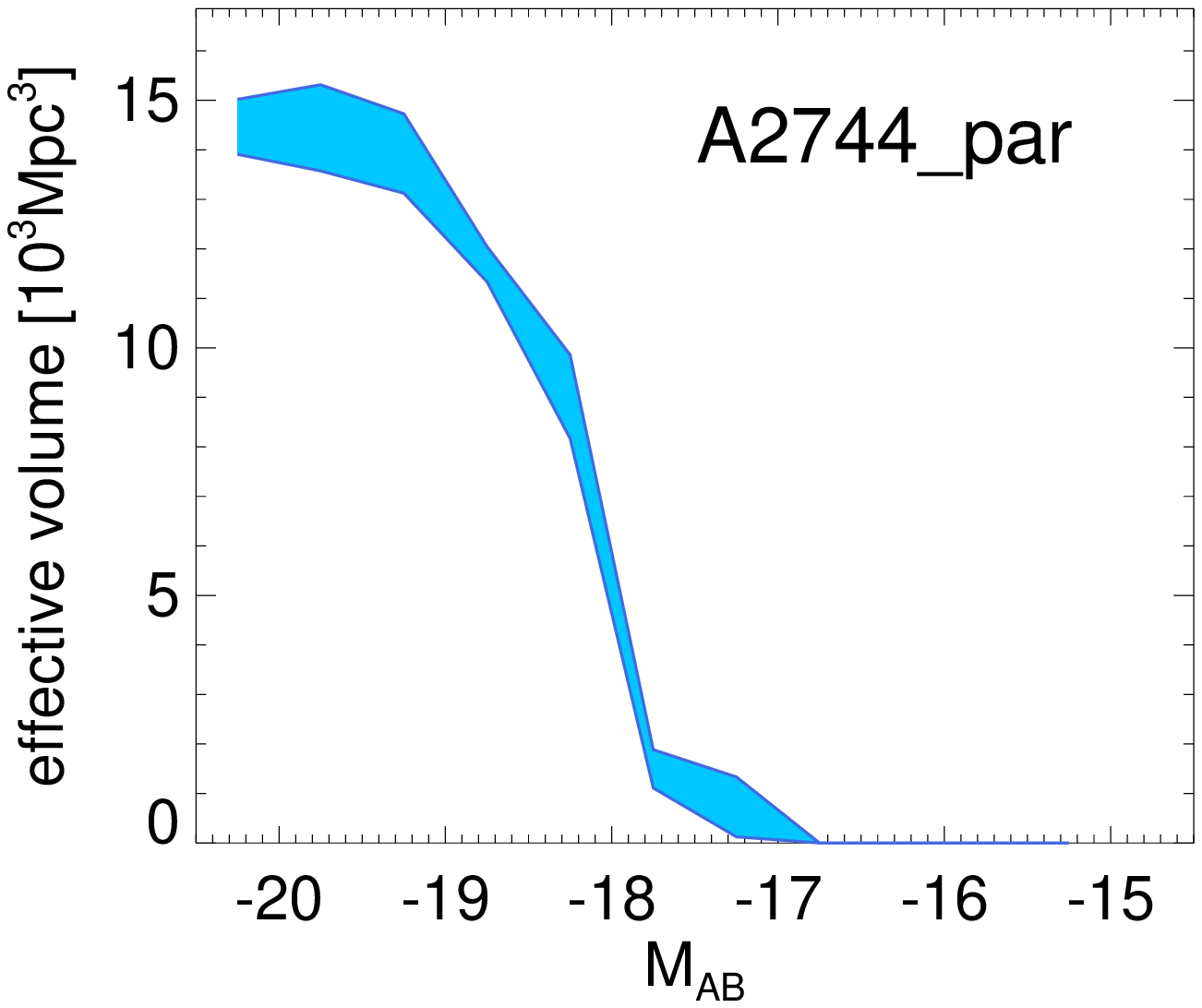} \\
    \includegraphics[width=4cm]{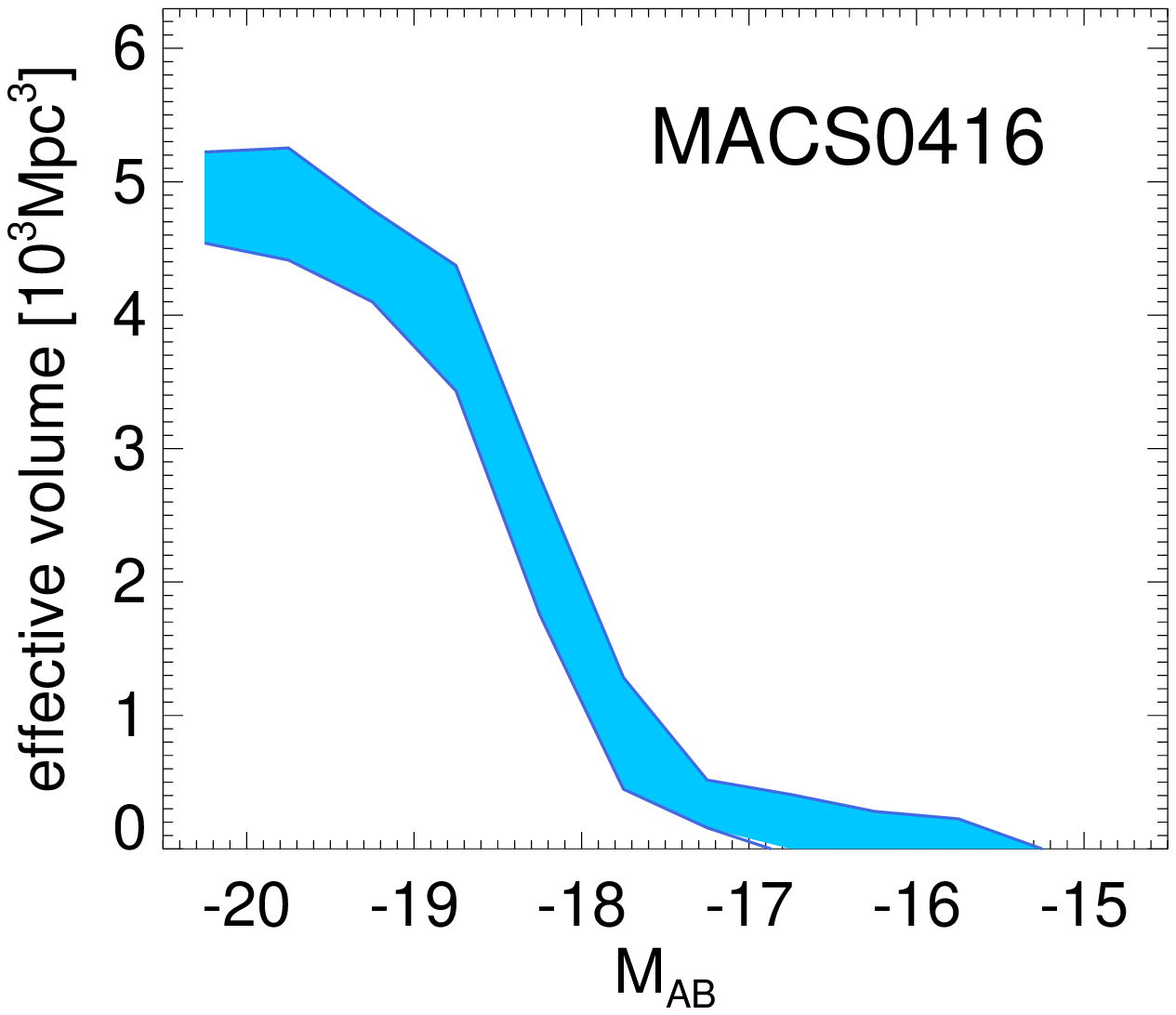} 
      \includegraphics[width=4cm]{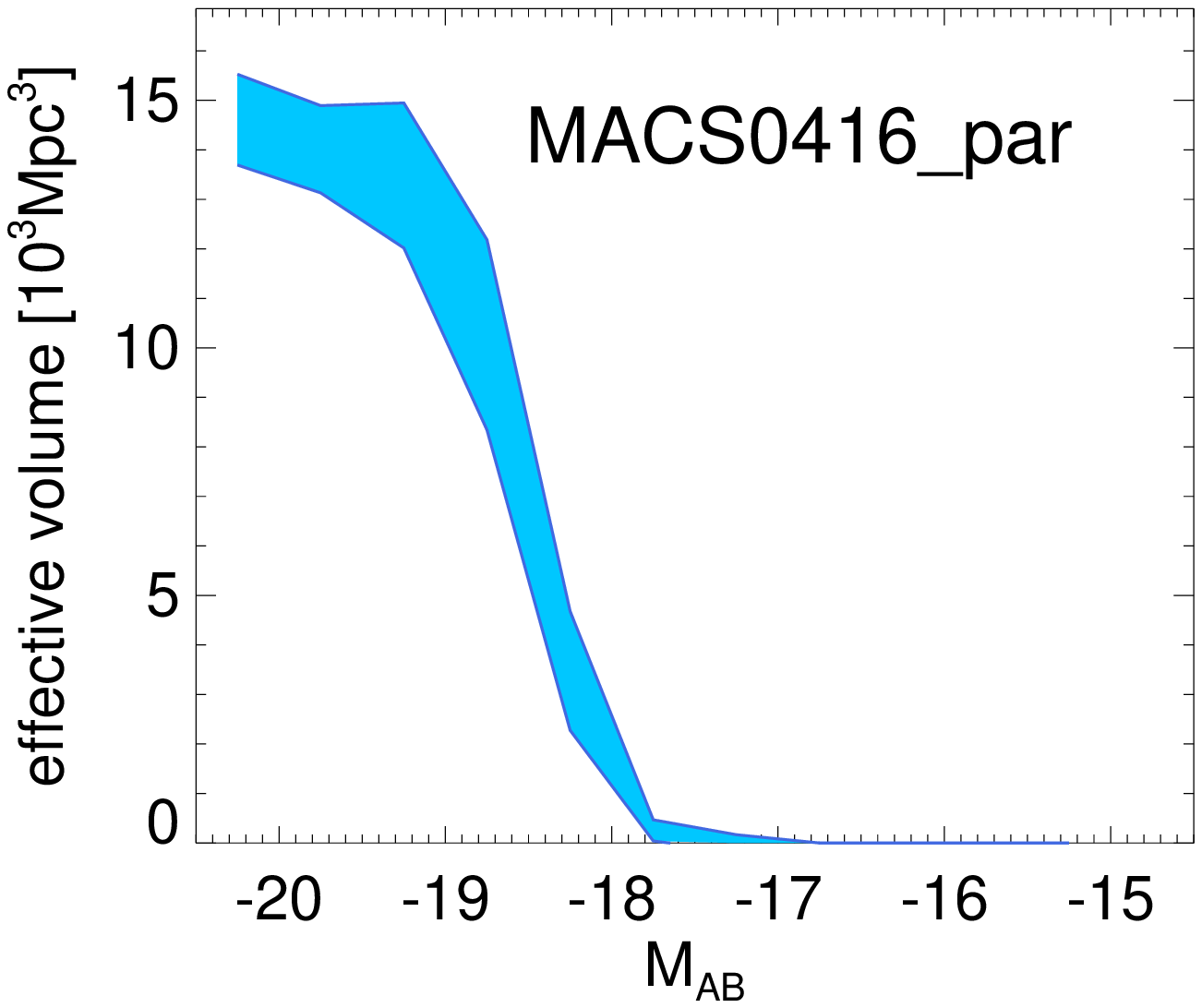} \\
       \includegraphics[width=4cm]{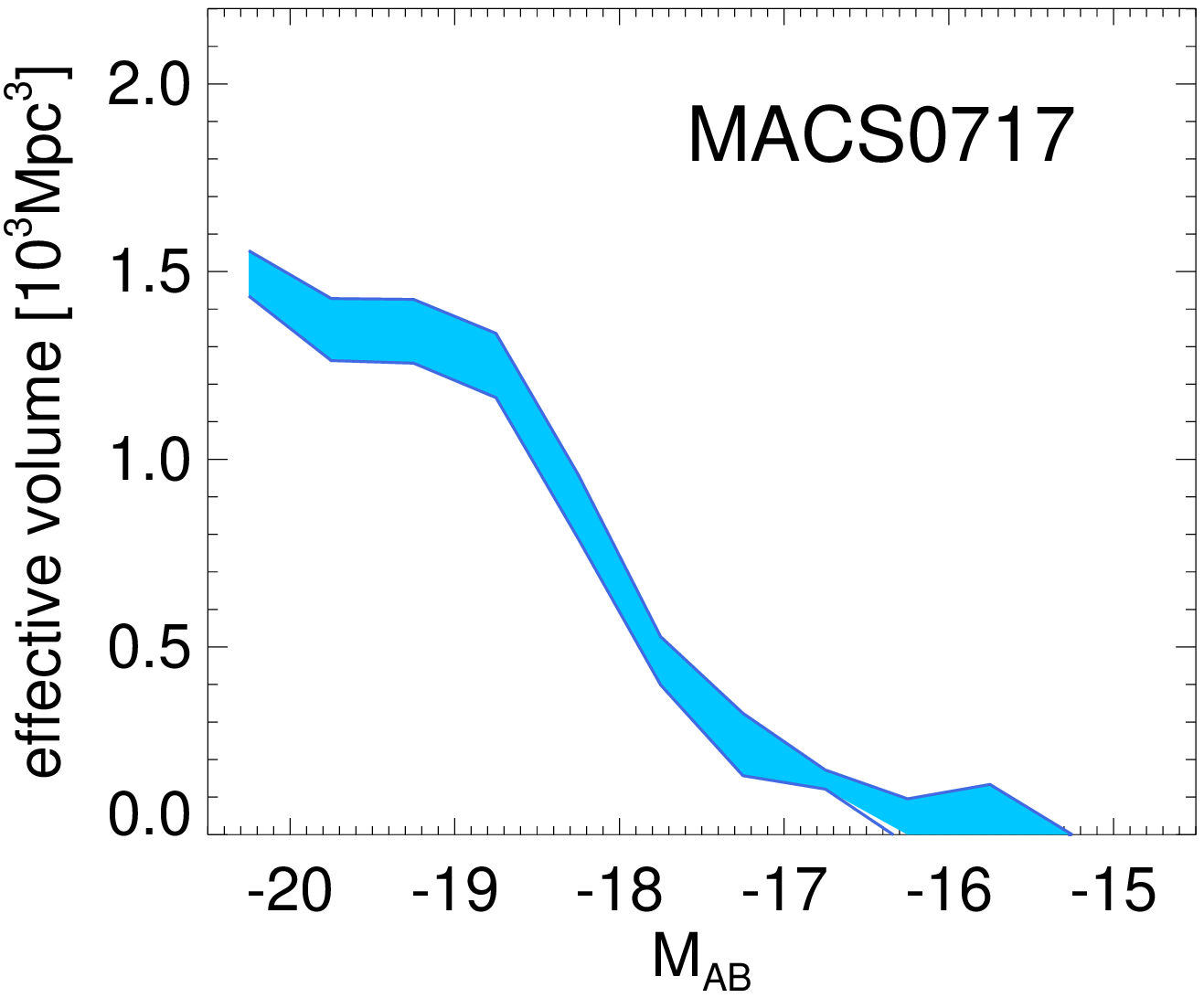} 
      \includegraphics[width=4cm]{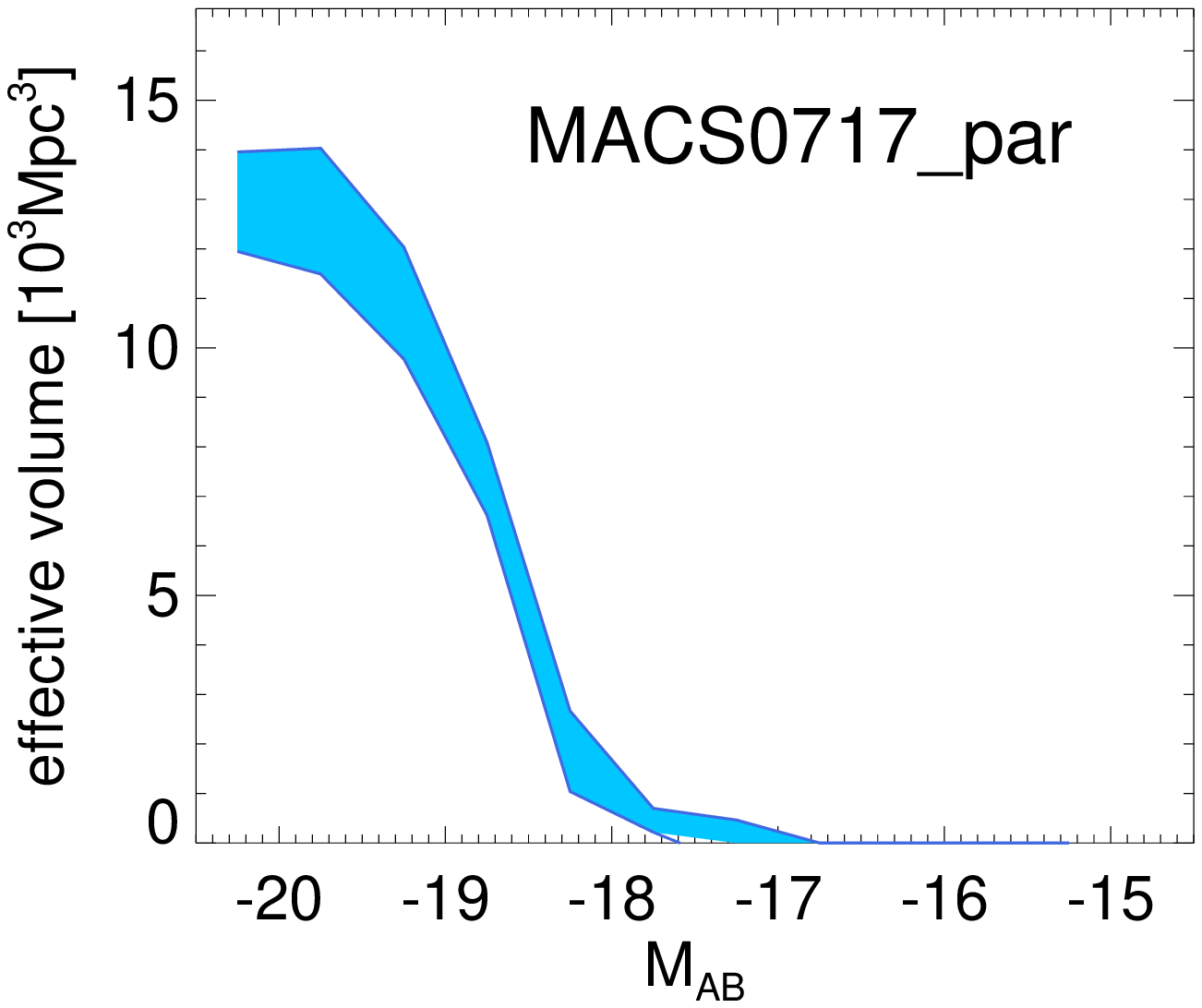} 
   \caption{The effective survey volume as a function of absolute UV magnitude for the redshift $z \sim 7$ sample. Each curve is based on our completeness simulations for each cluster or parallel field. The total area for the cluster fields is corrected for lensing effects whereas the blank field area is based on the full WFC3 field of view (cf. text for a detailed explanation). The blue colored region represents the 95\% confidence intervals of the completeness estimate.} 
   \label{fig:comp}
\end{figure}

Following \citet{atek15} we run extensive Monte Carlo simulations to assess the completeness level as a function of the intrinsic magnitude. A total of 10,000 galaxies were simulated for each of the six fields. We take the galaxy profile into account by creating two samples of exponential disks and de Vaucouleur shapes \citep{ferguson04,hathi08}. These galaxy profiles are then distorted by gravitational lensing according to our mass model. As for galaxy sizes, we adopt a log-normal distribution with a mean half light radius (hlr) of 0.15\arcsec\ and sigma=0.07\arcsec. For consistency with previous results \citep{ferguson04, bouwens04, hathi08, oesch10, grazian11, grazian12, huang13, oesch14}, the distribution is based on the sizes derived from spectroscopically confirmed LBG samples at $z\sim4$ \citep{vanzella09}, {\em while accounting for a redshift evolution of the intrinsic physical size of galaxies with a factor of ($1+z$)$^{-1}$}. {\em HST} observations of dropout galaxies at $z > 6$ redshifts also reveal small sizes of less than 0.3\arcsec\ \citep{mosleh12,ono13} and smaller for galaxies fainter than $M_{UV}-21$ mag. More recently \citet{kawamata15} measured the size of lensed dropout galaxies in the HFF cluster A2744 and report similar results. {Small sizes around 0.1\arcsec have also been observed in lensed galaxies by \citet{coe13,zitrin14}}. In addition, based on the results of \citep{huang13}, we adopt a size luminosity relation $r \propto L^{\beta}$, with $\beta = 0.25$, for our simulations \citep[see also][]{mosleh12, kawamata15}.

After assigning random redshifts in the range [5.5-7.5], we simulate galaxy magnitudes in each {\em HST} band using star-forming SED templates from \citep{kinney96}. In this step, we also assign random intrinsic absolute magnitude (in the rest-frame UV) in the range $M_{UV}=$[-14,-24] mag. We then include 10 simulated galaxies in the actual images of each band for a total of thousand images. This is where the cluster and parallel fields are treated differently in the simulations. In the cluster fields, galaxies are not included directly in the images but are simulated in the source plane. They are lensed into the image plane using the corresponding mass model. This way we ensure that all the lensing effects including magnification, shape distortion, position relative to the critical line are fully taken into account. Finally, for all the images, we run the same procedure used to select high-$z$ galaxies and determine the completeness function, which represents the fraction of the original galaxies recovered in our selection as a function of the parameters described above. This completeness function is in turn incorporated in the computation of the effective volume in each magnitude bin following the equation:

\begin{eqnarray}
V_{eff} = \int_{0}^{\infty}  \int_{\mu > \mu_{min}} \frac{dV_{com}}{dz}~ f(z,m,\mu) ~d\Omega(\mu,z) ~dz
\end{eqnarray}

where $\mu_{min}$ is the minimum amplification factor $\mu_{min}$ required to detect a galaxy with a given apparent magnitude $m$ and $V_{com}$ is the comoving volume. $f(z,m,\mu)$ is the completeness function that depends on the redshift $z$, apparent magnitude $m$, and amplification factor $\mu$, and $d\Omega(\mu)$ is the area element in the source plane, which is function of magnification and redshift.

Figure \ref{fig:comp} presents the results of our effective volume estimates in each field, marginalized over the intrinsic absolute UV magnitude. The extent of each filled region represents the 68\% confidence intervals. We can clearly see the importance of gravitational lensing in extending the survey depth to fainter galaxies. While the blank fields completeness drop abruptly before $M_{UV}=-18$ mag, it becomes shallower in cluster fields and extends down to $M_{UV}=-15$ mag, although at a level of 10\% or less.

\begin{figure*}[!htbp]
   \centering
   \includegraphics[width=5.5cm]{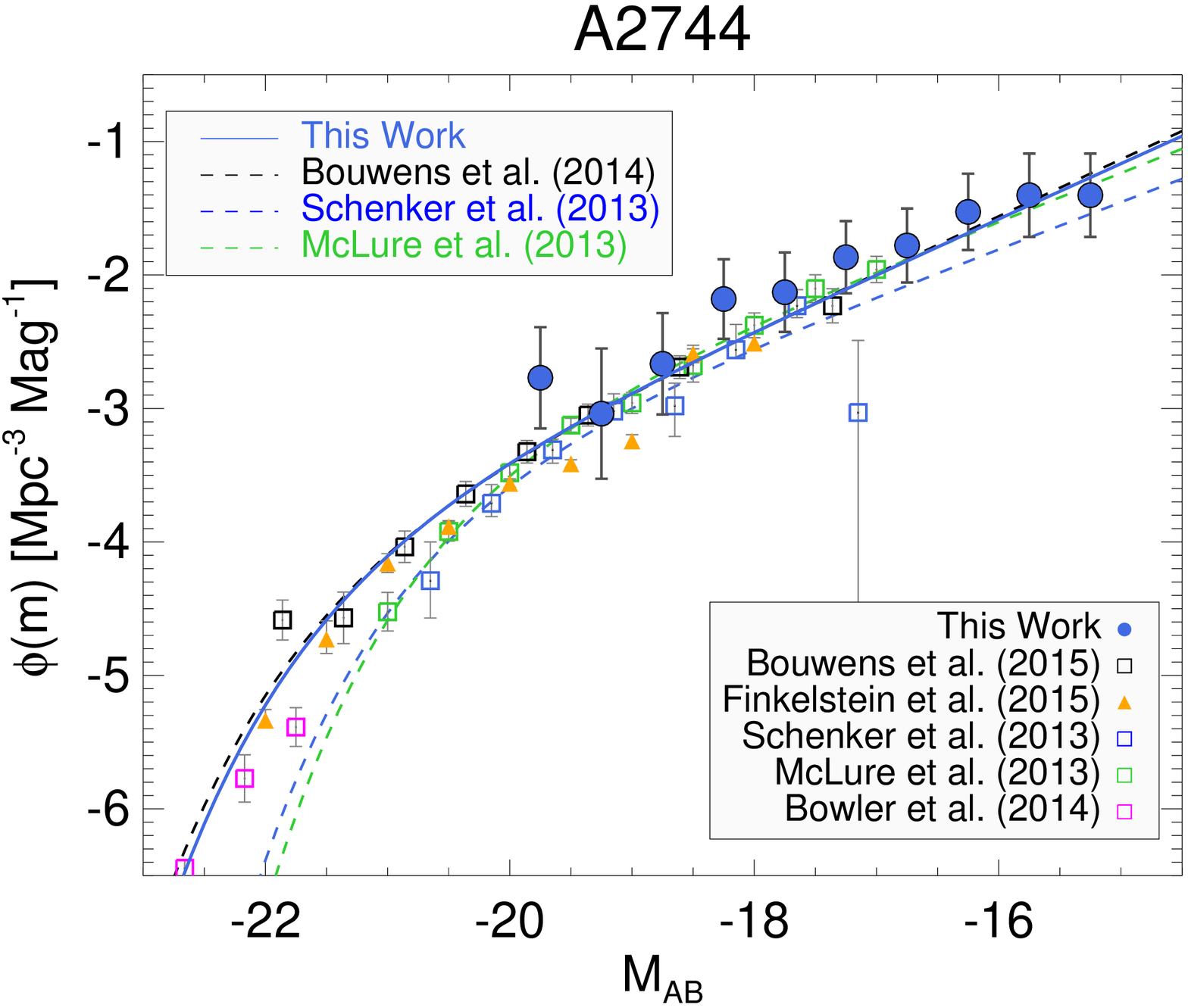} 
    \includegraphics[width=5.5cm]{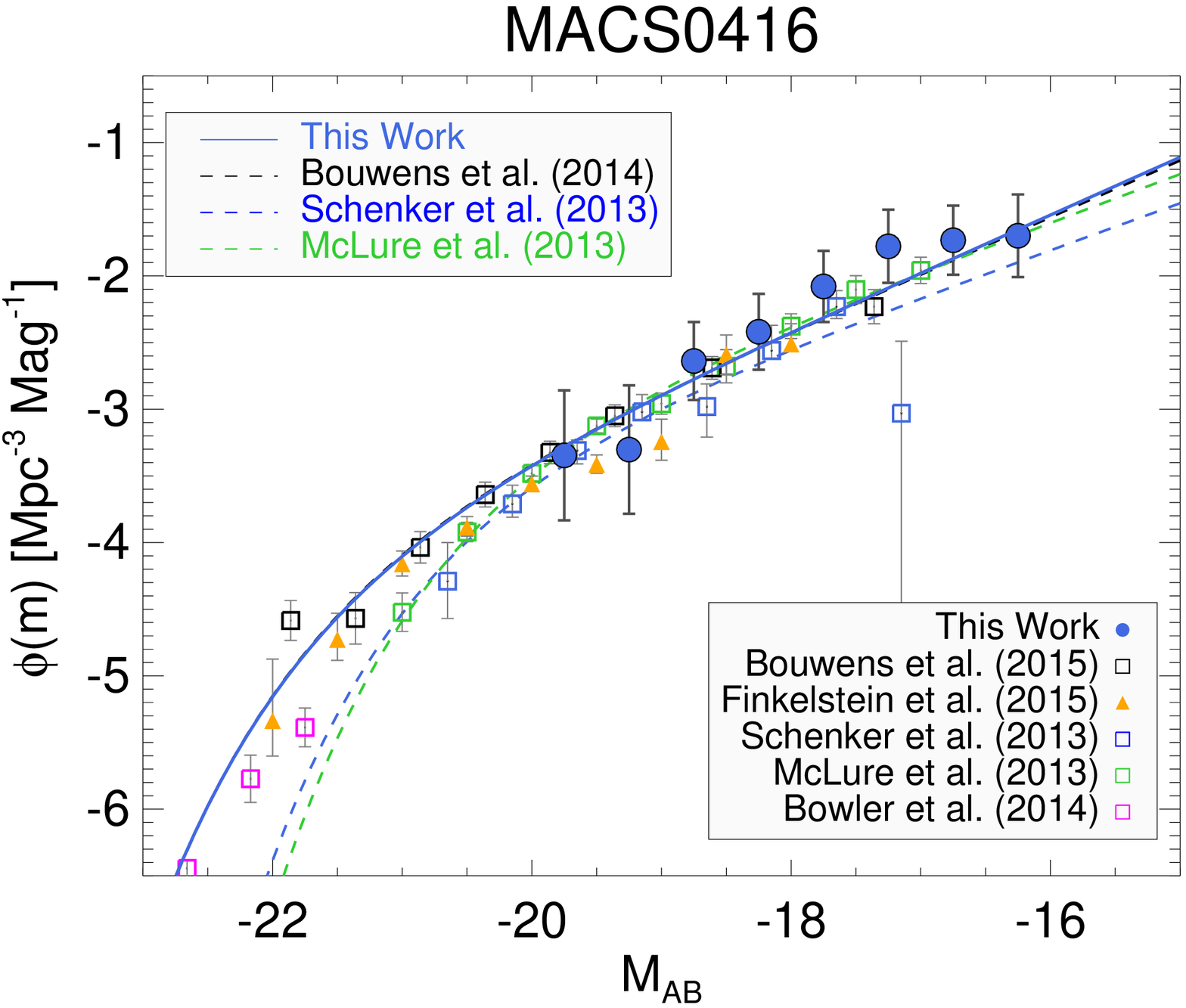} 
     \includegraphics[width=5.5cm]{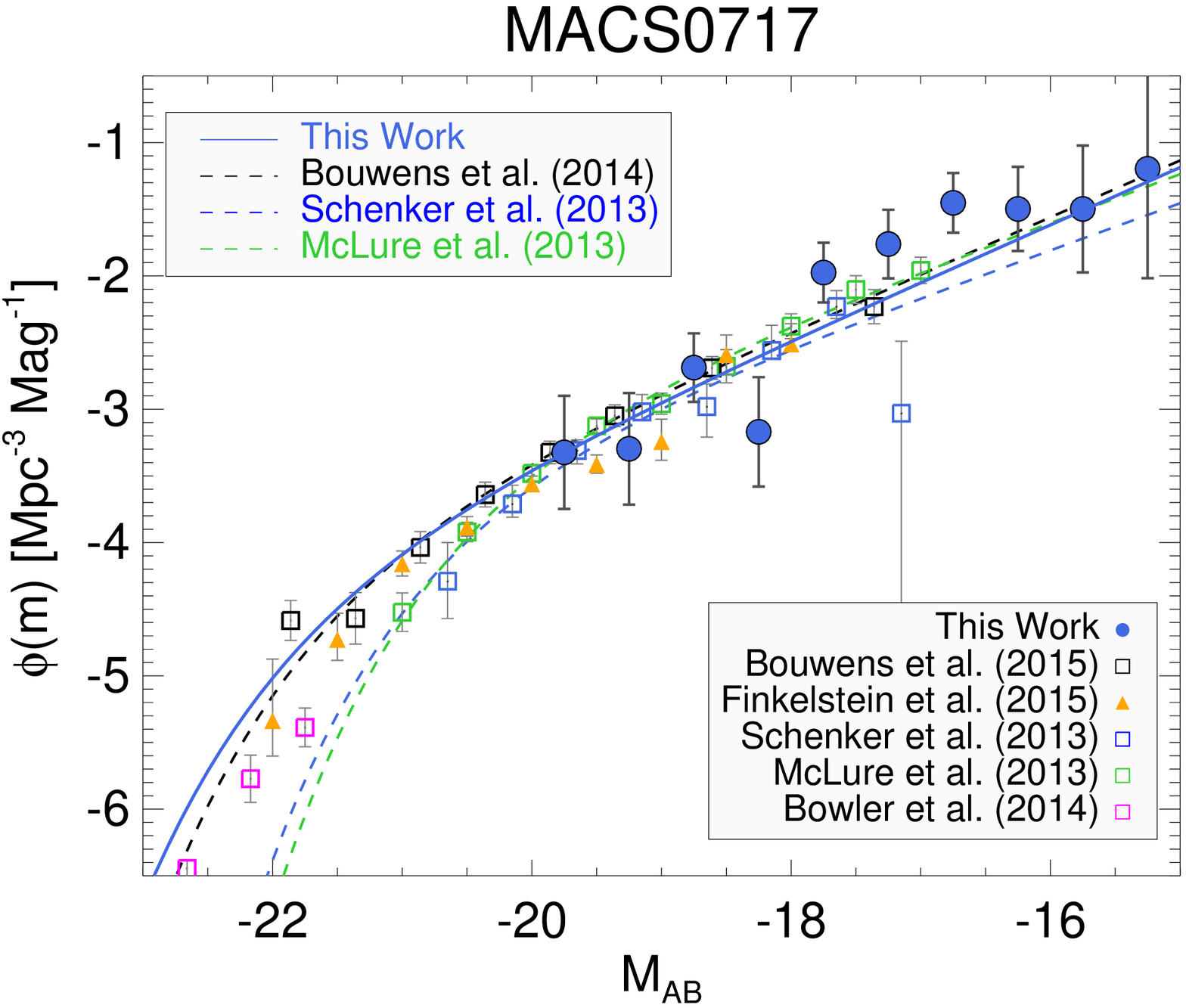} \\
      \includegraphics[width=5.5cm]{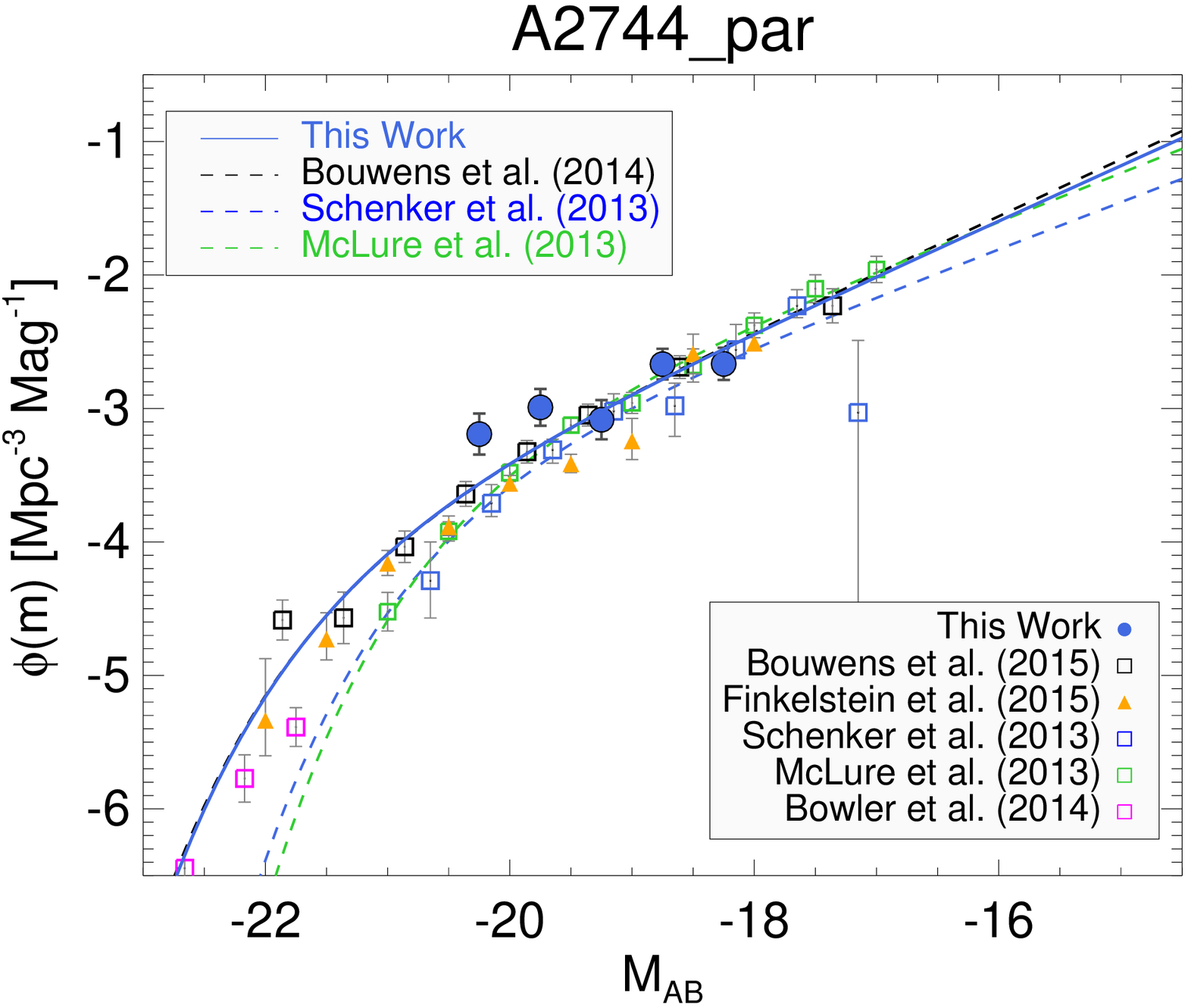} 
       \includegraphics[width=5.5cm]{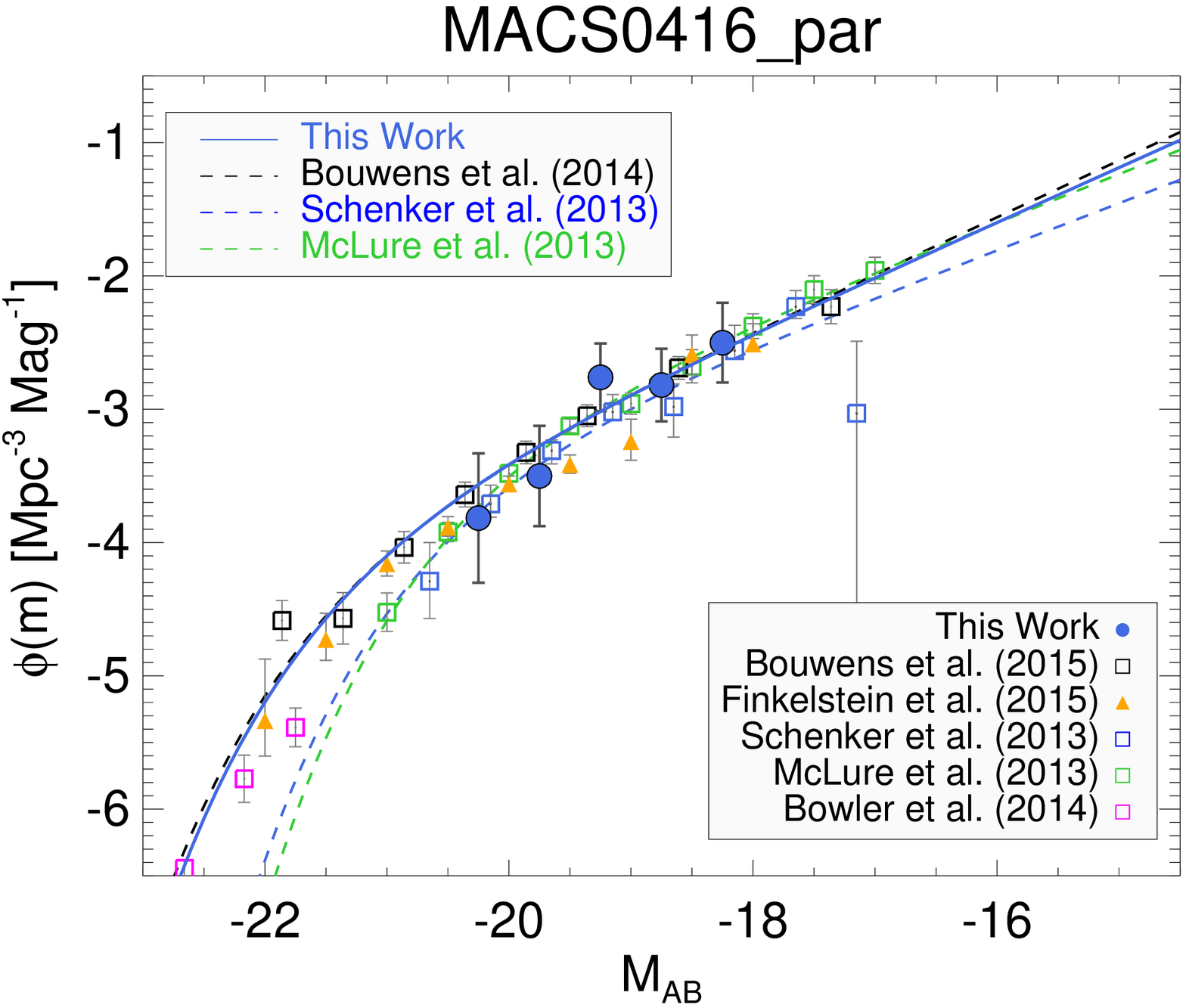} 
        \includegraphics[width=5.5cm]{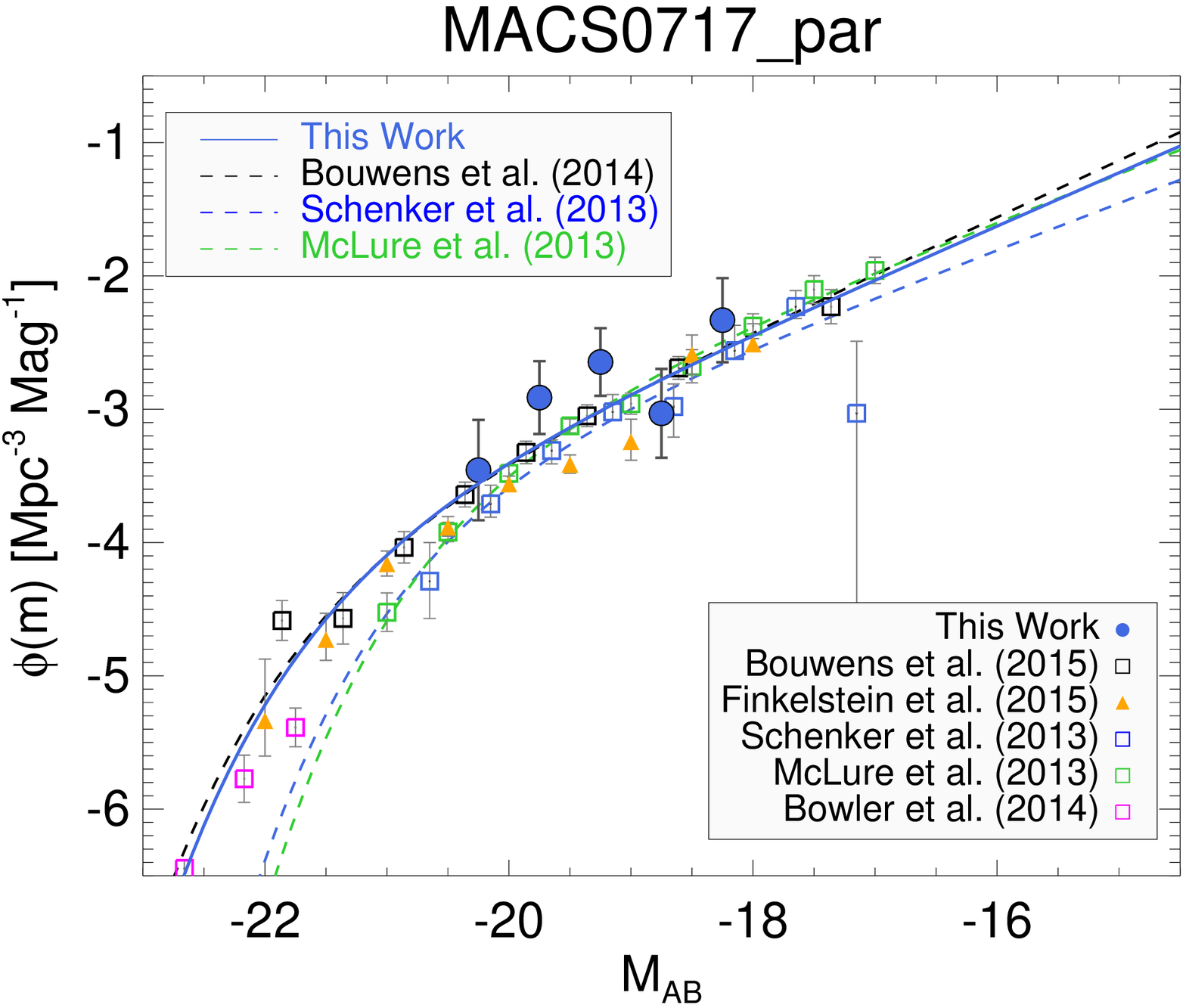} 
   \caption{UV luminosity function at $z \sim 7$ computed individually in each field. The blue circles represent our determination with 1-$\sigma$ uncertainties while the blue solid curve is our best Schechter function fit to the LF. We compare our results to previous literature results in blank fields. The black squares and dashed curve are from a compilation of {\em HST} legacy fields by \citet{bouwens14}. We also show the LF determination of \citet[][blue squares and dashed curve]{schenker13b} and \citet[][green squares and dashed curve]{mclure13} derived in the UDF12 field. We also include data points on the bright-end of the LF [magenta squares] from a wide area survey by \citet{bowler14}.}
   \label{fig:lf_ind}
\end{figure*}

\begin{table}
\centering
\caption{The best fit $z \sim 7$ Schechter parameters in each individual field}
\label{tab:lf}
\begin{tabular}[c]{l c c c}
\hline
\hline
Field &  $M_{UV}^\star$ & $\alpha$ & $\log_{10} \phi^\star$ \\
 & [AB mag] & & [Mpc$^{-3}$] \\
\hline
A2744        & $-20.92 \pm 0.64$	& $-2.03 \pm 0.13$	& $-3.56 \pm 0.45$	 \\
MACS0416 &  $-21.04 \pm 1.10$	& $-2.07 \pm 0.16$	& $-3.67 \pm 0.59$	 \\
MACS0717 &  $-21.13 \pm 1.52$	& $-2.02 \pm 0.22$	& $-3.73 \pm 0.83$	 \\
A2744 par   &  $-20.99 \pm 1.05$	& $-2.03 \pm 0.19$	& $-3.62 \pm 0.55$	 \\
MACS0416 par &  $-20.94 \pm 1.19$	& $-2.02 \pm 0.20$	& $-3.59 \pm 0.70$	 \\
MACS0717 par &  $-20.89 \pm 1.10$	& $-2.00 \pm 0.21$	& $-3.54 \pm 0.62$	 \\
\hline
\end{tabular}
\end{table}

\subsection{The UV LF at $z=6-7$}
\label{sec:lf_z7}

Using the derived effective volume as a function of absolute magnitude we compute the UV luminosity function following the equation:

\begin{eqnarray}
\phi(M)dM = \frac{N_{i}}{V_{eff}(M_{i})},
\end{eqnarray}

where $N_{i}$ and $M_{i}$ are the number of galaxies and the absolute magnitude, respectively, in each magnitude bin. Following most of the studies in the literature, we choose a bin size of 0.5 mag whereas the magnitude varies from one field to another. The individual LF determinations in each field are presented in Figure \ref{fig:lf_ind}. The top panels show the results in the cluster fields while the bottom panels are for parallel fields. The LF extends to $M_{UV}=-18.25$ in blank fields, whereas it reaches a magnitude of $M_{UV}=-15.25$ in cluster fields thanks to the lensing magnification. This gain can already be seen in the completeness function results (see Fig. \ref{fig:comp}) which reaches fainter magnitudes in cluster fields. Figure \ref{fig:amp} also shows the distribution of the amplification factor for the dropout samples selected behind the three lensing clusters. The flux amplification ranges essentially from $\mu \sim 1.25$ to 75, with a median value of 5.2 , 3.9, and 4.9 in A2744, MACS0416, and MACS0717, respectively.

In addition to the sample contamination discussed in Sect. \ref{sec:contamination}, several sources contribute to the uncertainties of the LF data points. The mass model errors, affecting the magnification and the survey volume (or the source plane area $\Omega$), are propagated into the LF determination in the case of cluster fields. We also include Poisson errors, and the cosmic variance estimate based on the recent results of \citet{robertson15} for both cluster and parallel fields. Finally, uncertainties from incompleteness simulations, as seen in Fig. \ref{fig:comp} are incorporated in the final LF. The bright-end is dominated by cosmic variance errors and small number counts, while the faint-end error bars reflect mostly the large incompleteness uncertainties at these faint magnitudes and small statistics.

\begin{figure}[!htbp]
   \centering
   \includegraphics[width=9cm]{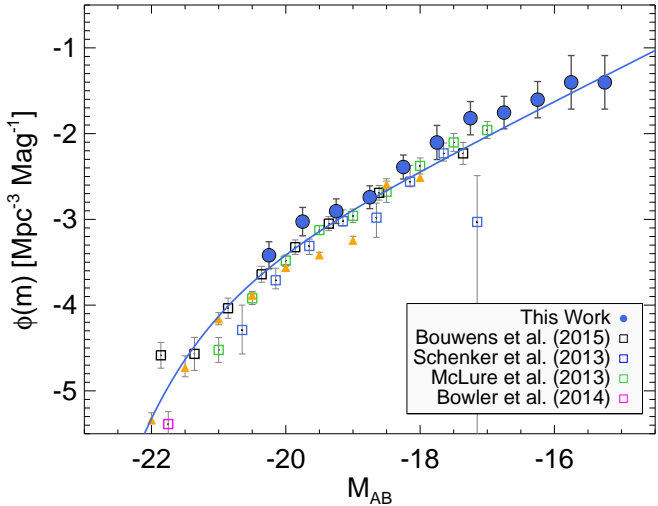} 
   \caption{The combined constraints from the clusters and parallel fields on the UV luminosity function at $z \sim 7$. The color code for the data points and the best fit Schechter function are the same as in Fig. \ref{fig:lf_ind}. }
   \label{fig:lf_combine}
\end{figure}

\begin{figure*}[htbp]
   \centering
   \includegraphics[width=6cm]{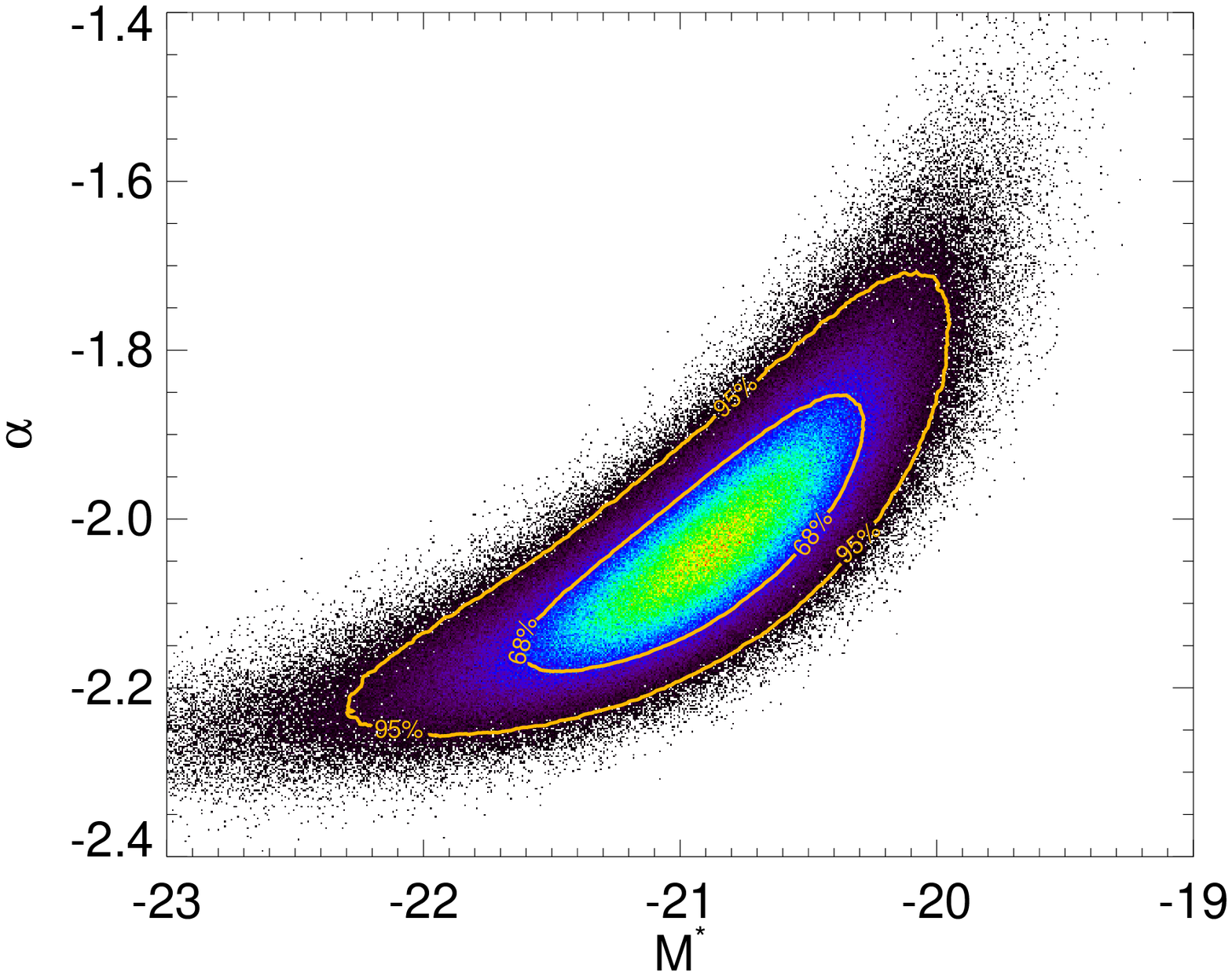} 
   \hspace{-0.5cm}
    \includegraphics[width=6cm]{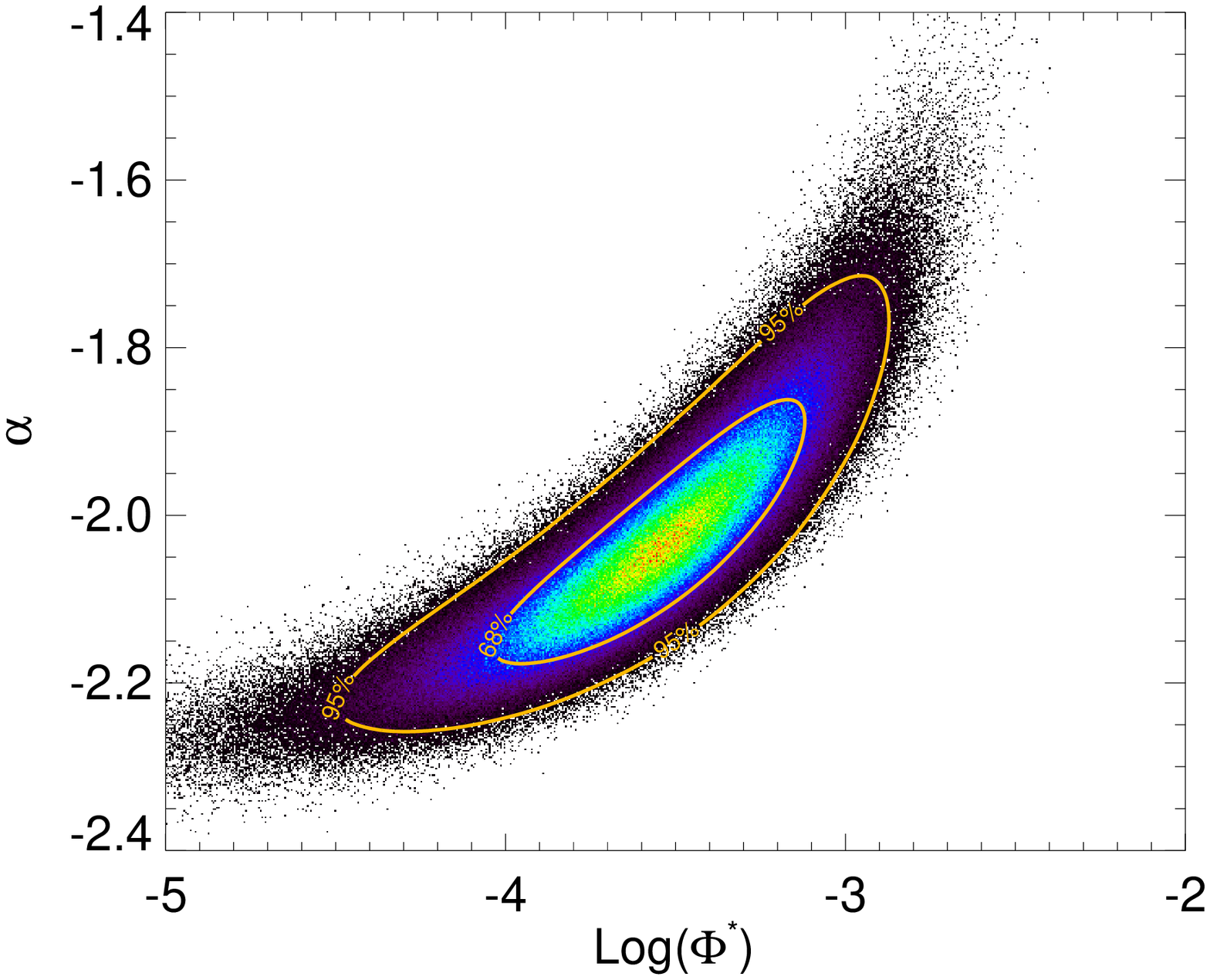} 
    \hspace{-0.5cm}
     \includegraphics[width=6cm]{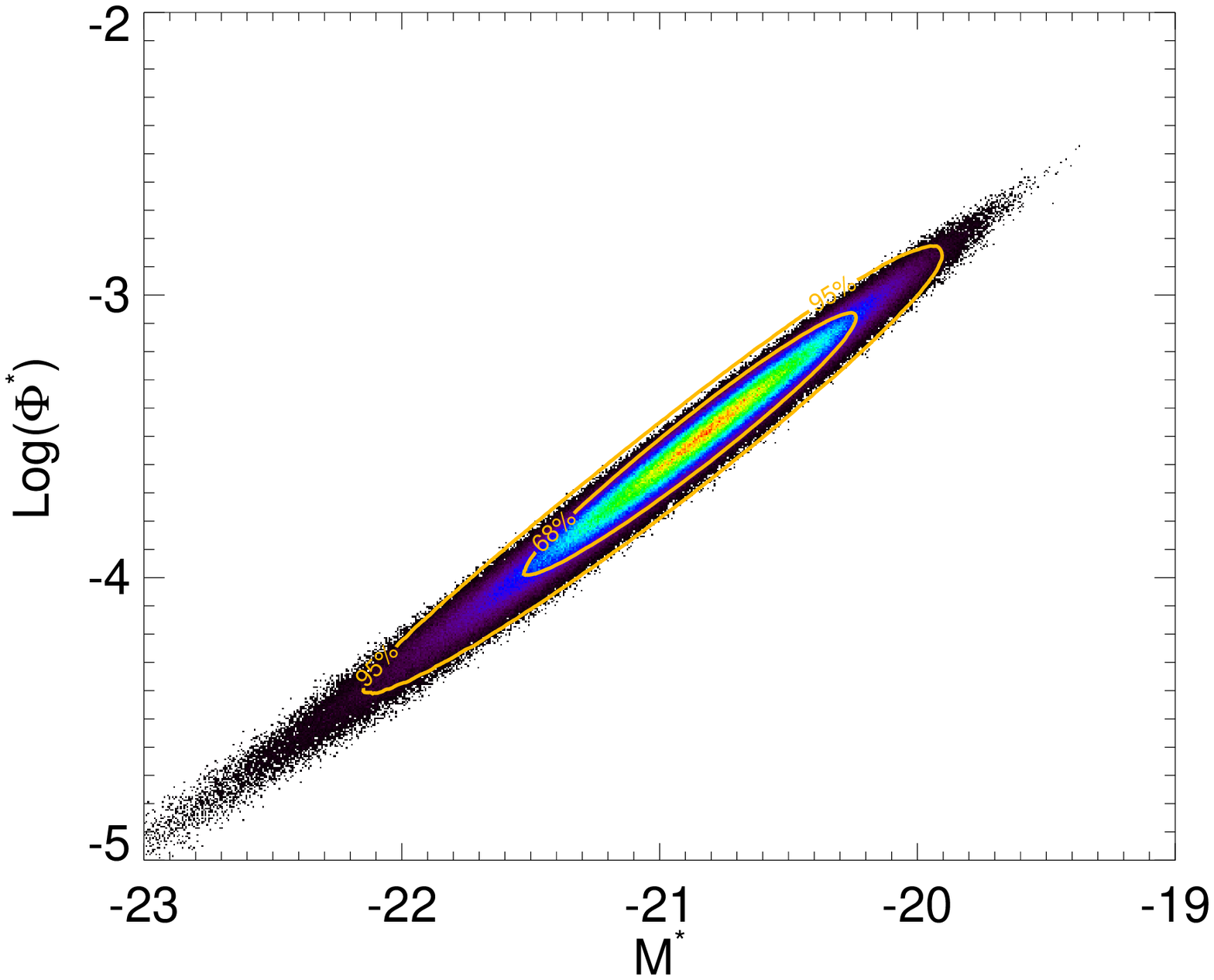} 
   \caption{Likelihood analysis of the Schechter parameters for the UV LF at $z \sim 7$: the faint-end slope $\alpha$, the characteristic $M^{\star}_{UV}$ and Log($\phi^{star}$). The density plots show the result of MCMC simulations marginalized over two parameters in each of the three panels.The orange curves represent the 68\% and 95\% confidence contours.}
   \label{fig:lf_lh}
\end{figure*}

We determine the shape of the rest-frame UV LF by fitting a Schechter function \citep{schechter76} to our data, which has been extensively used to describe the galaxy UV LF across a wide redshift range \citep[e.g.][]{bunker04,beckwith06,reddy09, bouwens06,mclure09,ouchi09,wilkins10,oesch12,willott13}. The Schechter function can be expressed in terms of absolute magnitudes as:

\begin{eqnarray}
\phi(M)=\frac{\rm{ln}(10)}{2.5} \phi^{\star}10^{0.4(\alpha+1)(M^{\star}-M)} exp(-10^{0.4(M^{\star}-M)}) \hspace{0.5cm}
\label{eq:lf}
\end{eqnarray}

We perform a Schechter fit to the LF in each field and find a faint-end slope $\alpha$ between -2.0 and -2.07 (cf. Eq. \ref{eq:lf}). We include in the fit the data points from \citet[][black squares]{bouwens14} to constrain the bright-end of the LF. The faintest bin is reached in A2744 at $M_{UV}=-15.25$. There is a hint of a shallower slope at those magnitudes that could be the result of large uncertainties in the completeness estimate, which is typically less then 10\% in this bin. As a matter of fact, a similar decline is observed in MACS0416 around $M_{UV}=-16.25$, whereas the slope is constantly steep at these magnitudes in A2744. It is clear that the constraints on the LF in MACS0717 are not as good as in the other cluster fields. This is likely the result of larger uncertainties on the lensing model due to the complex structure of the galaxy cluster. For instance, the model prediction for image position has an rms error of $\sim 1.9$ arcsec, while it is is about 0.7 to 0.8 arcsec in MACS0416 and A2744. These kind of deviations in the LF data points can also be observed in A2744 when using an old model based on pre-HFF observations that similarly yielded large uncertainties. Nonetheless, the overall shape of the LF remains consistent with the other clusters' results.

We now combine all the LF constraints from the lensed and parallel fields to compute the most robust UV LF at $z \sim 7$. The result is shown in Fig. \ref{fig:lf_combine}. Together with our data points, we show the most recent results from the literature from the blank fields as described in the legend. We ran MCMC simulations with $10^{6}$ realizations to find the best fit to the LF and estimate the uncertainties on the Schechter parameters. We find that the UV LF at $z \sim 7$ has a faint-end slope of $\alpha=-2.04_{-0.17}^{+0.13}$, a characteristic $M^{\star} = -20.89_{-0.72}^{+0.60}$ and Log($\phi^{\star}$)=$-3.54_{-0.45}^{+0.48}$. This is an excellent agreement with earlier results presented in \citet{atek15}, where we computed the UV LF in the HFF cluster A2744. In Figure \ref{fig:lf_lh}, we show the likelihood analysis of the Schechter parameters at $z \sim 7$, marginalized over two parameters at a time. Our results are in good agreement with the most recent results reported from the blank fields (cf. Table \ref{tab:lf}). In particular, the faint-end slope is close to the determination of \citet{finkelstein14} with $\alpha = -2.03_{-20}^{+21}$ and \citet{bouwens14} with $\alpha=-2.03\pm0.14$. {Our results show only a slightly steeper faint-end slope than the values reported in the Hubble extreme deep field \citep{schenker13,mclure13} with $\alpha=1.90_{-0.15}^{+0.14}$ and $\alpha=1.90_{-0.15}^{+0.14}$, respectively, and agree within the reported uncertainties. Also using the gravitational lensing of the first HFF cluster A2744, \citet{ishigaki15} find a slightly shallower slope of $\alpha = -1.94_{-0.10}^{+0.09}$, which still in agreement with our value within the errors.}. We also, performed a schechter fit while excluding data points from MACS0717 cluster, which yields very similar parameters: $\alpha=-2.03$, a characteristic $M^{\star} = -20.86$ and Log($\phi^{\star}$)=$-3.52$. Importantly, we note that the uncertainties on the faint-end slope decreases to $\sigma_{\alpha} \sim 0.1$.

\begin{table}
\centering
\caption{Combined constraints on the UV LF at $z \sim 7$}
\label{tab:obs} 
\begin{tabular}{lcc}
\hline
$M_{UV}$ & Log($\varphi$) & $\varphi_{err}$ \\ \hline
-20.25& -3.4184 & 0.1576 \\
-19.75 & -3.0263 & 0.1658 \\
-19.25 & -2.9044 & 0.1431 \\
-18.75 & -2.7418 & 0.1332 \\
-18.25 & -2.3896 & 0.1401 \\
-17.75 & -2.1032 & 0.1990 \\ 
-17.25 & -1.8201 &0.1940 \\
-16.75& -1.7548&  0.1893 \\
-16.25& -1.6044& 0.2117 \\
-15.75& -1.4012& 0.3123 \\
-15.25& -1.4012& 0.3122 \\ \hline
\end{tabular}
\end{table}

Several theoretical models and cosmological simulations produce predictions for the UV LF at high redshift. For instance, our LF results at $z \sim 7$ are in good agreement with the hydrodynamical simulations of \citet{jaacks12} with $M^{\star} = -20.82$ and Log($\phi^{\star}$)=$-3.74$, although they predict a steeper faint-end slope of $\alpha=-2.30$ down to similar lower magnitude limit of $M_{UV}=-15$. Theoretical models by \citet{mason15} find a closer slope of $\alpha=-1.95\pm0.17$ down to a magnitude limit of $M_{UV} = -12$, corresponding to a halo mass of $10^{9}$ \msol. Similarly, \citet{kimm14} find a theoretical faint magnitude limit of $M_{UV}=-13$ with a faint-end slope of $\alpha=-1.9$. Semi-analytical models of \citet{dayal14} also find a steep faint-end slope of $\alpha \sim -2.02$ at $z \sim 7$ in good agreement with our observations.

\begin{table}
\centering
\caption{Comparison of the best fit $z \sim 7$ Schechter parameters}
\label{tab:lf}
\begin{tabular}[c]{l c c c}
\hline
\hline
Reference &  $M_{UV}^\star$ & $\alpha$ & $\log_{10} \phi^\star$ \\
 &  [AB mag]& & [Mpc$^{-3}$] \\
\hline
This work             	& $-20.89_{-0.72}^{+0.60}$	& $-2.04_{-0.13}^{+0.17}$	& $-3.54_{-0.45}^{+0.48}$	 \\
\citet{atek15} $^{a}$	& $-20.90^{+0.90}_{-0.73}$	& $-2.01^{+0.20}_{-0.28}$	& $-3.55_{-0.57}^{+0.57}$	 \\
\citet{ishigaki15} $^{a}$& $-20.45^{+0.1}_{-0.2}$	& $-1.94^{+0.09}_{-0.10}$	& $-3.30^{+0.10}_{-0.20}$	 \\
\citet{bouwens14}	& $-21.04 \pm 0.26$	& $-2.06 \pm 0.12$	& $-3.65^{+0.27}_{-0.17}$	 \\
\citet{finkelstein14}	& $-21.03^{+0.37}_{-0.50}$	& $-2.03_{-0.20}^{+0.21}$	& $-3.80^{+0.41}_{-0.26}$	 \\
\cite{mclure13}        & $-19.90^{+0.23}_{-0.28}$	   & $-1.90^{+0.14}_{-0.15}$ 	& $-3.35^{+0.28}_{-0.45}$ \\ 
\hline
\multicolumn{4}{l}{ $^\textrm{a}$ Using the first HFF cluster A2744}\\
\end{tabular}
\end{table}

\subsection{The UV LF at $z\sim8$}
\label{sec:lf_z8}

\begin{figure}[!htbp]
   \centering
   \includegraphics[width=9cm]{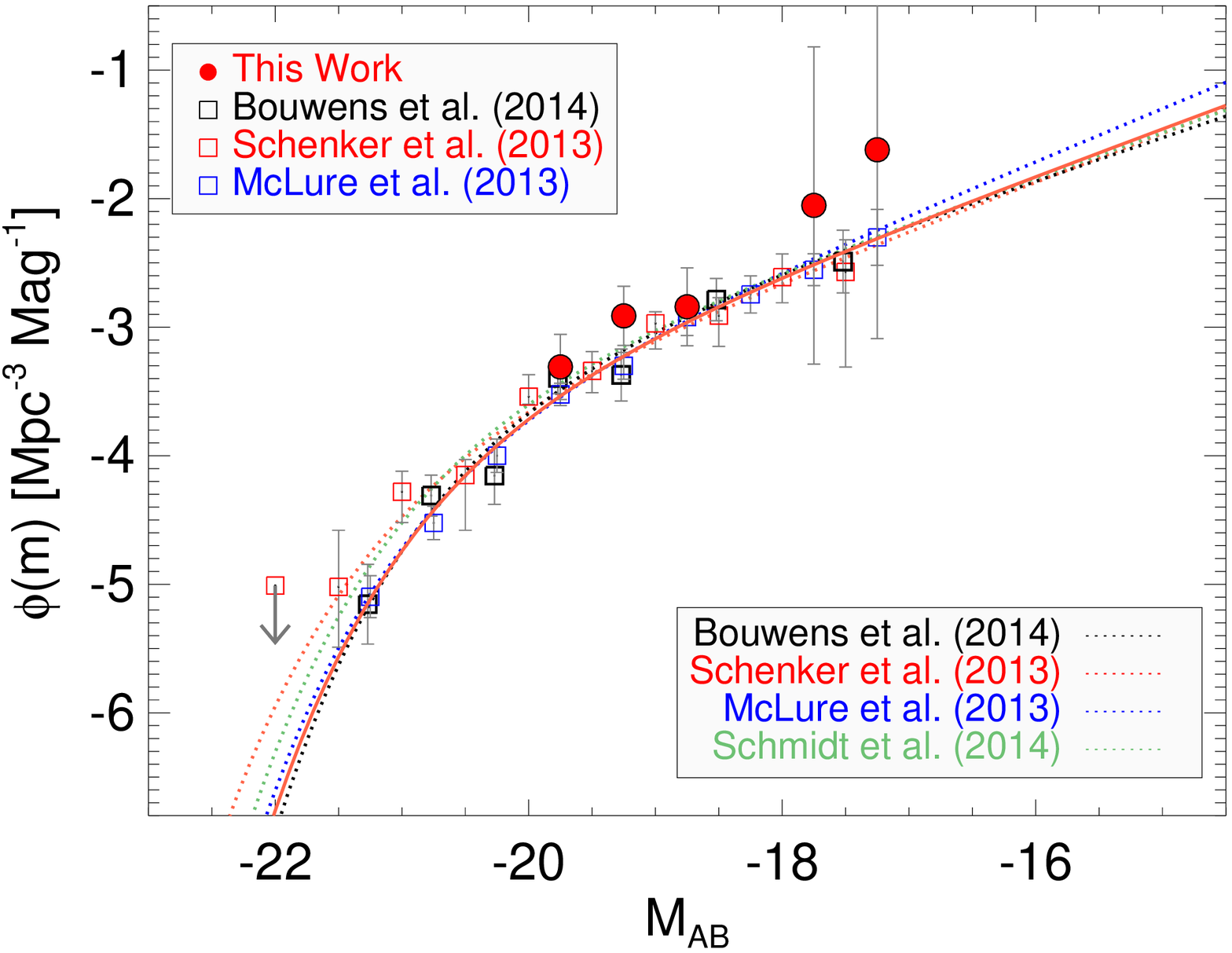} 
   \caption{The combined constraints from the clusters and parallel fields on the UV luminosity function at $z \sim 8$. The red circles represent our LF determination while the red squares are taken from \citet{schenker13}, the black squares from \citet{bouwens14}, and the blue squares from \citet{mclure13}. The green curve is the unbinned UV LF of \citet{schmidt14}. The best Schechter fits from the same literature results are also shown with dotted lines (cf. legend in the inset).}
   \label{fig:lf_combine_z8}
\end{figure}

Regarding the redshift $z \sim 8$ LF, we followed the same procedure used for the $z \sim 7$ LF. However, in lensing clusters, the survey volume at $z \sim 8$ is much more significantly reduced than at $z \sim 7$. In the extreme case of MACS0717, the total survey volume is about 460 Mpc$^3$, whereas it reaches 3600 Mpc$^{3}$ in A2744. Therefore, we expect lower galaxy number counts at $z \sim 8$ for high magnification values. In total, we detect only two galaxies with intrinsic magnitudes fainter than $M_{UV}=-18$, where the uncertainties on the LF estimate are very large (see Fig. \ref{fig:lf_combine_z8}). Unlike the $z \sim 7$ LF, we do not have strong constraints on the faint-end part of the LF at $z \sim 8$. At brighter magnitudes the LF is better constrained thanks to the addition of the parallel fields and appears in agreement with previous results in the literature. As discussed in \citet{atek15}, the redshift 8 galaxy selection in A2744 clearly shows an overdensity \citep[see also][]{zheng14,ishigaki15}, which translates into a large excess in the UV LF. Here we decided not to exclude the entire A2744 field in the combined LF to avoid introducing a well known bias due to cosmic variance \citep[see also][]{ishigaki15b}.

\subsection{Implications for cosmic reionization}

\begin{figure*}[htbp]
   \centering
   \includegraphics[width=15cm]{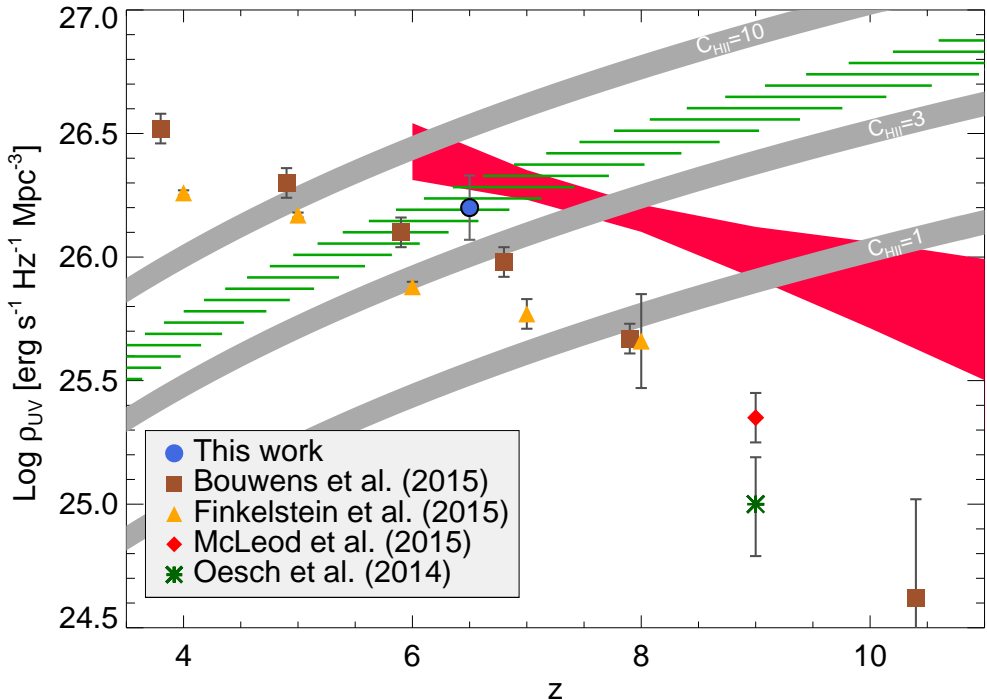} 
   \caption{The evolution of the UV luminosity density of galaxies with redshift. Our determination (blue circle) is based on the integration of the UV LF down to $M_{UV}=-15$. The brown squares represent the most recent results of \citet{bouwens14} who integrate the UV LF down to $M_{UV}=-17$. The orange triangles are the result of \citep{finkelstein14} who integrate to conservative limit of $M_{UV}=-18$. The red diamond and the green asterisk are the UV luminosity densities derived at $z \sim 9$ by \citet{mcleod15} and \citet{oesch14}, respectively. Both studies integrate the UV LF down to an observational limit of $M_{UV}=-17.7$, which correspond to a star formation rate of SFR$_{lim}=0.7$ \msol\ yr$^{-1}$. All the error bars indicate 1-$\sigma$ uncertainties. The shaded grey region show denote the UV luminosity density based on the ionizing emissivity \citep{madau99} required to maintain the IGM ionized at a certain redshift for and escape fraction of ionizing radiation of $f_{esc}\sim 20\%$ and three different values for the clumping factor C$_{\rm H\textsc{ii}}=1$, 3, and 10. The red shaded region shows the 68\% confidence interval for the evolution of the galaxy UV luminosity density based on constraints on the ionizing emissivity of \citet{bouwens15}. The green hatched region corresponds to the UV luminosity density required to maintain the IGM ionized for a clumping factor of C$_{\rm H\textsc{ii}}=3$ and escape fraction in the range $f_{esc}=10-15$\%, which encompasses our value at $z \sim 7$.}
   \label{fig:reionization}
\end{figure*}

In this study, we reliably extend the UV LF to unprecedented depth, and put strong constraints on the faint-end slope down to $M_{UV}=-15.25$. The most important result is that the faint-end slope remains very steep down to a luminosity of 0.005$L^{\star}$, the characteristic UV luminosity at $z \sim 7$. Also, thanks to the combination of six fields, we have significantly reduced the uncertainties on the faint-end slope to about 5\%. With these strong constraints in hand, we can now integrate the UV LF to derive the UV luminosity density at $z \sim 7$. The main advantage of these observations is the ability to set the integration lower limit to $M_{UV} = -15$. Unlike previous results that extrapolate the UV LF to lower magnitude, we use an observational constraint to estimate the ultraviolet photon budget from galaxies. For instance, based on the UV LF results in blank fields, \citet{bouwens14} report a total UV luminosity density of Log($\rho_{UV})=25.98\pm0.06$ erg s$^{-1}$ Hz$^{-1}$ Mpc$^{-3}$ and \citet{finkelstein14} a value of Log($\rho_{UV})=25.77\pm0.06$ erg s$^{-1}$ Hz$^{-1}$ Mpc$^{-3}$ at $z \sim 7$. Both values are computed down to $M_{UV}=-17$ AB, which is the limiting magnitude of their observations. Here we compute the UV luminosity density down to $M_{UV} = -15$ and find Log($\rho_{UV}$)=$26.2 \pm 0.13$ erg s$^{-1}$ Hz$^{-1}$ Mpc$^{-3}$. 

In order to determine whether the UV luminosity produced by galaxies is sufficient to reionize the IGM, one needs to estimate additional parameters, for which observational constraints remain challenging. First, the conversion factor from the UV luminosity to ionizing radiation $\xi_{ion}$ (ergs$^{-1}$ Hz) depends on the star-formation history of galaxies. The value of $\xi_{ion}$ is generally constrained using stellar population models of early galaxies and the observed UV slope $\beta$ of $z > 6$ galaxies \citep{bolton07,kuhlen12,robertson15}. Once the ionizing photon production is determined we need to estimate its escape fraction $f_{esc}$ from galaxies to ionize the IGM. Direct observational constraints of $f_{esc}$ are difficult, especially at $z > 6$  because of the high opacity of the intervening hydrogen residuals on the line of sight. Many studies and deep surveys were dedicated to the search of ionizing continuum (Lyman continuum, LyC) escape from $z < 4$ galaxies. Very few detections were reported however and show very low escape fractions relative to the UV radiation, of the order of few percent \citep{shapley06, iwata09, siana10, nestor13}. More recently, \citet{debarros15} reported a spectroscopic detection of Lyman continuum emission in a $z \sim 3.2$ galaxy with a relative escape fraction\footnote{the relative escape fraction the ratio between the fraction of escaping Lyman continuum photons and the fraction of escaping photons at 1500 \AA\ \citep{steidel01}} of $f_{esc} \sim 65\%$.

In Figure \ref{fig:reionization}, we show our determination of the UV luminosity density together with the most recent results in the literature as a function of redshift including \citet{bouwens14} and \citet{finkelstein14} at $z=4-8$, \citet{bouwens14} at $z=10$, \citet{mcleod15} and \citet{oesch14} at $z \sim 9$. While their values are based on the integration of the UV LF down to a magnitude limit between $M_{UV}=-18$ and $M_{UV}=-17$, our integration limit is two magnitudes fainter at $M_{UV}=-15$. There is a significant difference between the results of \citet{bouwens14} and \citet{finkelstein14} due to the fact that the UV LF in the former study extends one magnitude deeper than the latter. \citet{finkelstein14} did not make use of the full IR data available in the HUDF, which also explains the larger uncertainties in their faint-end slope constraints. The UV luminosity density of galaxies at $z \sim 7$ determined in the present work is clearly larger than previous determinations, owing to a steep faint-end slope and a very faint integration limit. We can now assess whether this UV production is sufficient to ionize the IGM. We use the photon emission rate per unit cosmological
comoving volume required to maintain reionization at a given redshift determined in \citep{madau99}:

\begin{eqnarray}
\dot{N}(z) = ( 10^{51.2} s^{-1} Mpc^{-3} ) C_{30} \left(\frac{1+z}{6}\right)^{3} \left(\frac{\Omega_{b}h^{2}}{0.02}\right)^{2}
\end{eqnarray}

where $C_{30}$ is the clumping factor $C_{\rm H\textsc{ii}} = \langle n_{\rm H\textsc{ii}}^{2}  \rangle  /   \langle n_{\rm H\textsc{ii}} \rangle^{2}$ normalized to $C_{\rm H\textsc{ii}}=30$, and $n_{\rm H\textsc{ii}}$ is the mean comoving hydrogen density in the Universe. This is the minimum value for which the ionizing emission balances the recombination rate. We show in Fig. \ref{fig:reionization} this limit, converted to UV luminosity density \citep[cf.][]{madau99, bolton07} assuming an escape fraction of the ionizing radiation of $f_{esc}=20\%$ and three different values for the clumping factor. Assuming a standard value of $C_{H\rm{\textsc{ii}}}=3$ for the clumping factor of the IGM \citep{pawlik09,finlator12} our determination of galaxy UV luminosity density at $z \sim 7$ is sufficient to maintain reionization. We also show that an escape fraction as low as $f_{esc}=10-15$\% (green hatched region) is already sufficient to ionize the IGM at $z \sim 7$. At the same redshift, the ionizing emissivity constraints of \citet{bouwens15} are close to our UV density constraints.

With current observations we are not able to put better constraints on the faint-end of the UV LF at $z \sim 8$, hence on the ionizing emissivity of galaxies at this redshift. The conclusions from the deep blank fields are still affected by large uncertainties that prevent any strong claims regarding the contribution of galaxies to the ionizing budget at $z \sim 8$ \citep{bouwens14,finkelstein14}. Future {\em HST} observations of the remaining HFF clusters might add better constrains on the UV luminosity density at $z \sim 8$ with additional highly magnified faint galaxies.

\section{Summary}
\label{sec:summary}

Combining {\em HST} observations of lensing clusters and parallel fields of the Hubble Frontier Fields program, we computed the galaxy UV luminosity function between $z=6$ and $z=8$. We assembled a large sample of about 250 galaxy candidates in this redshift range using the Lyman break photometric selection. In the cluster cores, we corrected the deep {\em HST} images for intra-cluster light and bright cluster galaxies light using a median filtering before object detection, while performing the photometry in the original images. This technique helps the detection of faint galaxies contaminated by cluster light but do not increase significantly the number of galaxies in the cluster center because of the very small volume probed in those regions that have very high magnification.

\begin{itemize}
\item Using the latest lensing models produced by the CATS team for the three clusters A2744 \citep{jauzac15}, MACS0416 \citep{jauzac14b}, and MACS0717 (Limousin et al. 2015), we have performed completeness simulations in the source plane that take into account all lensing effects. Thanks to the lensing magnification, the completeness function extends down to fainter magnitudes in the cluster fields than in the parallels. On the other hand, the total survey volume goes from $\sim 4$ arcmin$^{2}$ in the parallel fields down to 0.6-1 arcmin$^{2}$ in lensing fields.   

\item We computed the UV luminosity function for all individual fields and for the combined sample at $z \sim 7$. The lensing magnification allow us to extend the LF down to an absolute UV magnitude limit of $M_{UV}=-15.25$, which is more than two magnitudes deeper than any study the deep blank fields. Most importantly, we show that the faint-end slope remains very steep at $\alpha = -2.04^{+0.13}_{-0.17}$ at such faint intrinsic luminosity. When excluding MACS0717 cluster, the uncertainties on the faint-end slope decrease to $\sigma_{\alpha} \sim 0.1$. The best Schechter fit yields a characteristic magnitude of $M^{\star} = -20.89_{-0.72}^{+0.60}$ and Log($\phi^{\star}$)=$-3.54_{-0.45}^{+0.48}$. This is in good agreement with most of the recent results in blank fields \citep{bouwens14,finkelstein14}, which were limited to $M_{UV}=-17.5$ and early results of \citet{atek15}.

\item Our determination of the UV LF at $z \sim 8$ does not reach beyond $M_{UV} = -18$ because we detect only two galaxies in the two fainter magnitude bins. Albeit with large uncertainties due to small number statistics and incompleteness uncertainties, the currently determined LF points confirm the results of previous studies.

\item On observational grounds, we integrate the UV LF down to a magnitude limit of $M_{UV}=-15$ and find Log($\rho_{UV})=26.2 \pm 0.13$ erg s$^{-1}$ Hz$^{-1}$ Mpc$^{-3}$. Assuming standard values for the ionizing conversion factor $\xi_{ion}$ and the clumping factor $C_{H\rm{\textsc{ii}}}=3$, the ionizing budget of galaxies would be sufficient to maintain the IGM ionized by $z \sim 7$ provided the ionizing escape fraction from galaxies is greater than 10\%. Future observations of the HFF program will allow us to decrease the uncertainties on the UV luminosity density at $z \sim 7$ and perhaps improve the constraints on the UV luminosity density at $z \sim 8$, which are mostly based on blank field observations.

\end{itemize}

With the help of gravitational lensing we have produced the best constraints currently available on the UV luminosity function and the contribution of galaxies to the IGM reionization at $z \sim 7$. We clearly demonstrate here the great potential and the feasibility of peering into the early universe through these cosmic telescopes. The remaining HFF clusters will certainly help improve even more our constraints on the UV LF at $z \sim 7$ and in particular at $z \sim 8$ to better assess the role of galaxies in the cosmic reionization process. Our results also show the great promise of future programs targeting lensing fields and pave the way to observing programs with the {\em James Webb Space Telescope} or the {\em Wide-Field Infrared Survey Telescope} ({\em WFIRST}), which are scheduled for launch in the near future.

\acknowledgments

We want to thank the STScI and the HFF team for their efforts in obtaining and reducing the {\em HST} data. HA and JPK are supported by the European Research Council (ERC) advanced grant ``Light on the Dark'' (LIDA). JR acknowledges support from the ERC starting grant CALENDS. MJ acknowledges support from the Leverhulme Trust (grant number PLP-2011-003) and Science and Technology Facilities Council (grant number ST/L00075X/1). PN acknowledges support from NSF theory grant AST-1044455 and a theory grant from Space Telescope Science Institute HST-AR1214401.A. ML acknowledges support from CNRS. 

\bibliographystyle{apj}
\bibliography{references.bib}

\begin{thebibliography}{129}
\expandafter\ifx\csname natexlab\endcsname\relax\def\natexlab#1{#1}\fi

\bibitem[{{Abell} {et~al.}(1989){Abell}, {Corwin}, \& {Olowin}}]{abell89}
{Abell}, G.~O., {Corwin}, Jr., H.~G., \& {Olowin}, R.~P. 1989, \apjs, 70, 1

\bibitem[{{Arnouts} {et~al.}(2005){Arnouts}, {Schiminovich}, {Ilbert},
  {Tresse}, {Milliard}, {Treyer}, {Bardelli}, {Budavari}, {Wyder}, {Zucca}, {Le
  F{\`e}vre}, {Martin}, {Vettolani}, {Adami}, {Arnaboldi}, {Barlow}, {Bianchi},
  {Bolzonella}, {Bottini}, {Byun}, {Cappi}, {Charlot}, {Contini}, {Donas},
  {Forster}, {Foucaud}, {Franzetti}, {Friedman}, {Garilli}, {Gavignaud},
  {Guzzo}, {Heckman}, {Hoopes}, {Iovino}, {Jelinsky}, {Le Brun}, {Lee},
  {Maccagni}, {Madore}, {Malina}, {Marano}, {Marinoni}, {McCracken}, {Mazure},
  {Meneux}, {Merighi}, {Morrissey}, {Neff}, {Paltani}, {Pell{\`o}}, {Picat},
  {Pollo}, {Pozzetti}, {Radovich}, {Rich}, {Scaramella}, {Scodeggio},
  {Seibert}, {Siegmund}, {Small}, {Szalay}, {Welsh}, {Xu}, {Zamorani}, \&
  {Zanichelli}}]{arnouts05}
{Arnouts}, S., {et~al.} 2005, \apjl, 619, L43

\bibitem[{{Atek} {et~al.}(2009){Atek}, {Schaerer}, \& {Kunth}}]{atek09b}
{Atek}, H., {Schaerer}, D., \& {Kunth}, D. 2009, \aap, 502, 791

\bibitem[{{Atek} {et~al.}(2011){Atek}, {Siana}, {Scarlata}, {Malkan},
  {McCarthy}, {Teplitz}, {Henry}, {Colbert}, {Bridge}, {Bunker}, {Dressler},
  {Fosbury}, {Hathi}, {Martin}, {Ross}, \& {Shim}}]{atek11}
{Atek}, H., {et~al.} 2011, \apj, 743, 121

\bibitem[{{Atek} {et~al.}(2014{\natexlab{a}}){Atek}, {Kneib}, {Pacifici},
  {Malkan}, {Charlot}, {Lee}, {Bedregal}, {Bunker}, {Colbert}, {Dressler},
  {Hathi}, {Lehnert}, {Martin}, {McCarthy}, {Rafelski}, {Ross}, {Siana}, \&
  {Teplitz}}]{atek14c}
---. 2014{\natexlab{a}}, \apj, 789, 96

\bibitem[{{Atek} {et~al.}(2014{\natexlab{b}}){Atek}, {Richard}, {Kneib},
  {Clement}, {Egami}, {Ebeling}, {Jauzac}, {Jullo}, {Laporte}, {Limousin}, \&
  {Natarajan}}]{atek14b}
---. 2014{\natexlab{b}}, \apj, 786, 60

\bibitem[{{Atek} {et~al.}(2015){Atek}, {Richard}, {Kneib}, {Jauzac},
  {Schaerer}, {Clement}, {Limousin}, {Jullo}, {Natarajan}, {Egami}, \&
  {Ebeling}}]{atek15}
---. 2015, \apj, 800, 18

\bibitem[{{Beckwith} {et~al.}(2006){Beckwith}, {Stiavelli}, {Koekemoer},
  {Caldwell}, {Ferguson}, {Hook}, {Lucas}, {Bergeron}, {Corbin}, {Jogee},
  {Panagia}, {Robberto}, {Royle}, {Somerville}, \& {Sosey}}]{beckwith06}
{Beckwith}, S.~V.~W., {et~al.} 2006, \aj, 132, 1729

\bibitem[{{Bennett} {et~al.}(2013){Bennett}, {Larson}, {Weiland}, {Jarosik},
  {Hinshaw}, {Odegard}, {Smith}, {Hill}, {Gold}, {Halpern}, {Komatsu}, {Nolta},
  {Page}, {Spergel}, {Wollack}, {Dunkley}, {Kogut}, {Limon}, {Meyer}, {Tucker},
  \& {Wright}}]{bennett13}
{Bennett}, C.~L., {et~al.} 2013, \apjs, 208, 20

\bibitem[{Bertin \& Arnouts(1996)}]{bertin96}
Bertin, E., \& Arnouts, S. 1996, Astronomy and Astrophysics Supplement Series,
  117, 393

\bibitem[{{Bolton} \& {Haehnelt}(2007)}]{bolton07}
{Bolton}, J.~S., \& {Haehnelt}, M.~G. 2007, \mnras, 382, 325

\bibitem[{{Bouwens} {et~al.}(2012){Bouwens}, {Bradley}, {Zitrin}, {Coe},
  {Franx}, {Zheng}, {Smit}, {Host}, {Postman}, {Moustakas}, {Labbe},
  {Carrasco}, {Molino}, {Donahue}, {Kelson}, {Meneghetti}, {Benitez}, {Lemze},
  {Umetsu}, {Broadhurst}, {Moustakas}, {Rosati}, {Bartelmann}, {Ford},
  {Graves}, {Grillo}, {Infante}, {Jiminez-Teja}, {Jouvel}, {Lahav}, {Maoz},
  {Medezinski}, {Melchior}, {Merten}, {Nonino}, {Ogaz}, \& {Seitz}}]{bouwens12}
{Bouwens}, R., {et~al.} 2012, ArXiv e-prints

\bibitem[{{Bouwens} {et~al.}(2004){Bouwens}, {Illingworth}, {Blakeslee},
  {Broadhurst}, \& {Franx}}]{bouwens04}
{Bouwens}, R.~J., {Illingworth}, G.~D., {Blakeslee}, J.~P., {Broadhurst},
  T.~J., \& {Franx}, M. 2004, \apjl, 611, L1

\bibitem[{{Bouwens} {et~al.}(2006){Bouwens}, {Illingworth}, {Blakeslee}, \&
  {Franx}}]{bouwens06}
{Bouwens}, R.~J., {Illingworth}, G.~D., {Blakeslee}, J.~P., \& {Franx}, M.
  2006, \apj, 653, 53

\bibitem[{{Bouwens} {et~al.}(2015{\natexlab{a}}){Bouwens}, {Illingworth},
  {Oesch}, {Caruana}, {Holwerda}, {Smit}, \& {Wilkins}}]{bouwens15}
{Bouwens}, R.~J., {Illingworth}, G.~D., {Oesch}, P.~A., {Caruana}, J.,
  {Holwerda}, B., {Smit}, R., \& {Wilkins}, S. 2015{\natexlab{a}}, ArXiv
  e-prints

\bibitem[{{Bouwens} {et~al.}(2015{\natexlab{b}}){Bouwens}, {Illingworth},
  {Oesch}, {Trenti}, {Labb{\'e}}, {Bradley}, {Carollo}, {van Dokkum},
  {Gonzalez}, {Holwerda}, {Franx}, {Spitler}, {Smit}, \& {Magee}}]{bouwens14}
{Bouwens}, R.~J., {et~al.} 2015{\natexlab{b}}, \apj, 803, 34

\bibitem[{{Bowler} {et~al.}(2014){Bowler}, {Dunlop}, {McLure}, {Rogers},
  {McCracken}, {Milvang-Jensen}, {Furusawa}, {Fynbo}, {Taniguchi}, {Afonso},
  {Bremer}, \& {Le F{\`e}vre}}]{bowler14}
{Bowler}, R.~A.~A., {et~al.} 2014, \mnras, 440, 2810

\bibitem[{{Brada{\v c}} {et~al.}(2005){Brada{\v c}}, {Schneider}, {Lombardi},
  \& {Erben}}]{bradac05}
{Brada{\v c}}, M., {Schneider}, P., {Lombardi}, M., \& {Erben}, T. 2005, \aap,
  437, 39

\bibitem[{{Bradley} {et~al.}(2012){Bradley}, {Bouwens}, {Zitrin}, {Smit},
  {Coe}, {Ford}, {Zheng}, {Illingworth}, {Ben{\'{\i}}tez}, \&
  {Broadhurst}}]{bradley12}
{Bradley}, L.~D., {et~al.} 2012, \apj, 747, 3

\bibitem[{{Bradley} {et~al.}(2014){Bradley}, {Zitrin}, {Coe}, {Bouwens},
  {Postman}, {Balestra}, {Grillo}, {Monna}, {Rosati}, {Seitz}, {Host}, {Lemze},
  {Moustakas}, {Moustakas}, {Shu}, {Zheng}, {Broadhurst}, {Carrasco}, {Jouvel},
  {Koekemoer}, {Medezinski}, {Meneghetti}, {Nonino}, {Smit}, {Umetsu},
  {Bartelmann}, {Ben{\'{\i}}tez}, {Donahue}, {Ford}, {Infante}, {Jimenez-Teja},
  {Kelson}, {Lahav}, {Maoz}, {Melchior}, {Merten}, \& {Molino}}]{bradley14}
---. 2014, \apj, 792, 76

\bibitem[{{Bunker} {et~al.}(2004){Bunker}, {Stanway}, {Ellis}, \&
  {McMahon}}]{bunker04}
{Bunker}, A.~J., {Stanway}, E.~R., {Ellis}, R.~S., \& {McMahon}, R.~G. 2004,
  \mnras, 355, 374

\bibitem[{{Bunker} {et~al.}(2010){Bunker}, {Wilkins}, {Ellis}, {Stark},
  {Lorenzoni}, {Chiu}, {Lacy}, {Jarvis}, \& {Hickey}}]{bunker10}
{Bunker}, A.~J., {et~al.} 2010, \mnras, 409, 855

\bibitem[{{Caruana} {et~al.}(2014){Caruana}, {Bunker}, {Wilkins}, {Stanway},
  {Lorenzoni}, {Jarvis}, \& {Ebert}}]{caruana14}
{Caruana}, J., {Bunker}, A.~J., {Wilkins}, S.~M., {Stanway}, E.~R.,
  {Lorenzoni}, S., {Jarvis}, M.~J., \& {Ebert}, H. 2014, \mnras, 443, 2831

\bibitem[{Casertano {et~al.}(2000)Casertano, de~Mello, Dickinson, Ferguson,
  Fruchter, Gonzalez-Lopezlira, Heyer, Hook, Levay, Lucas, \&
  et~al.}]{r_Hook_Levay_Lucas_et_al__2000}
Casertano, S., {et~al.} 2000, The Astronomical Journal, 120, 2747

\bibitem[{{Chabrier} {et~al.}(2000){Chabrier}, {Baraffe}, {Allard}, \&
  {Hauschildt}}]{chabrier00}
{Chabrier}, G., {Baraffe}, I., {Allard}, F., \& {Hauschildt}, P. 2000, \apj,
  542, 464

\bibitem[{{Chornock} {et~al.}(2014){Chornock}, {Berger}, {Fox}, {Fong},
  {Laskar}, \& {Roth}}]{chornock14}
{Chornock}, R., {Berger}, E., {Fox}, D.~B., {Fong}, W., {Laskar}, T., \&
  {Roth}, K.~C. 2014, ArXiv e-prints

\bibitem[{{Coe} {et~al.}(2015){Coe}, {Bradley}, \& {Zitrin}}]{coe15}
{Coe}, D., {Bradley}, L., \& {Zitrin}, A. 2015, \apj, 800, 84

\bibitem[{{Coe} {et~al.}(2013){Coe}, {Zitrin}, {Carrasco}, {Shu}, {Zheng},
  {Postman}, {Bradley}, {Koekemoer}, {Bouwens}, {Broadhurst}, {Monna}, {Host},
  {Moustakas}, {Ford}, {Moustakas}, {van der Wel}, {Donahue}, {Rodney},
  {Ben{\'{\i}}tez}, {Jouvel}, {Seitz}, {Kelson}, \& {Rosati}}]{coe13}
{Coe}, D., {et~al.} 2013, \apj, 762, 32

\bibitem[{{Coleman} {et~al.}(1980){Coleman}, {Wu}, \& {Weedman}}]{coleman80}
{Coleman}, G.~D., {Wu}, C.-C., \& {Weedman}, D.~W. 1980, \apjs, 43, 393

\bibitem[{{Dayal} {et~al.}(2014){Dayal}, {Ferrara}, {Dunlop}, \&
  {Pacucci}}]{dayal14}
{Dayal}, P., {Ferrara}, A., {Dunlop}, J.~S., \& {Pacucci}, F. 2014, \mnras,
  445, 2545

\bibitem[{{de Barros} {et~al.}(2015){de Barros}, {Vanzella}, {Amor{\'{\i}}n},
  {Castellano}, {Siana}, {Grazian}, {Suh}, {Balestra}, {Vignali}, {Verhamme},
  {Zamorani}, {Mignoli}, {Hasinger}, {Comastri}, {Pentericci},
  {P{\'e}rez-Montero}, {Fontana}, {Giavalisco}, \& {Gilli}}]{debarros15}
{de Barros}, S., {et~al.} 2015, ArXiv e-prints

\bibitem[{{Diego} {et~al.}(2014){Diego}, {Broadhurst}, {Molnar}, {Lam}, \&
  {Lim}}]{diego14}
{Diego}, J.~M., {Broadhurst}, T., {Molnar}, S.~M., {Lam}, D., \& {Lim}, J.
  2014, ArXiv e-prints

\bibitem[{{Duncan} \& {Conselice}(2015)}]{duncan15}
{Duncan}, K., \& {Conselice}, C.~J. 2015, \mnras, 451, 2030

\bibitem[{{Ebeling} {et~al.}(2007){Ebeling}, {Barrett}, {Donovan}, {Ma},
  {Edge}, \& {van Speybroeck}}]{ebeling07}
{Ebeling}, H., {Barrett}, E., {Donovan}, D., {Ma}, C.-J., {Edge}, A.~C., \&
  {van Speybroeck}, L. 2007, \apjl, 661, L33

\bibitem[{{Ebeling} {et~al.}(2001){Ebeling}, {Edge}, \& {Henry}}]{ebeling01}
{Ebeling}, H., {Edge}, A.~C., \& {Henry}, J.~P. 2001, \apj, 553, 668

\bibitem[{{Ebeling} {et~al.}(2010){Ebeling}, {Edge}, {Mantz}, {Barrett},
  {Henry}, {Ma}, \& {van Speybroeck}}]{ebeling10}
{Ebeling}, H., {Edge}, A.~C., {Mantz}, A., {Barrett}, E., {Henry}, J.~P., {Ma},
  C.~J., \& {van Speybroeck}, L. 2010, \mnras, 407, 83

\bibitem[{{Ebeling} {et~al.}(2014){Ebeling}, {Ma}, \& {Barrett}}]{ebeling14}
{Ebeling}, H., {Ma}, C.-J., \& {Barrett}, E. 2014, \apjs, 211, 21

\bibitem[{{El{\'{\i}}asd{\'o}ttir} {et~al.}(2007){El{\'{\i}}asd{\'o}ttir},
  {Limousin}, {Richard}, {Hjorth}, {Kneib}, {Natarajan}, {Pedersen}, {Jullo},
  \& {Paraficz}}]{eliasdottir07}
{El{\'{\i}}asd{\'o}ttir}, {\'A}., {et~al.} 2007, ArXiv e-prints

\bibitem[{{Fan} {et~al.}(2006){Fan}, {Strauss}, {Becker}, {White}, {Gunn},
  {Knapp}, {Richards}, {Schneider}, {Brinkmann}, \& {Fukugita}}]{fan06}
{Fan}, X., {et~al.} 2006, \aj, 132, 117

\bibitem[{{Ferguson} {et~al.}(2004){Ferguson}, {Dickinson}, {Giavalisco},
  {Kretchmer}, {Ravindranath}, {Idzi}, {Taylor}, {Conselice}, {Fall},
  {Gardner}, {Livio}, {Madau}, {Moustakas}, {Papovich}, {Somerville},
  {Spinrad}, \& {Stern}}]{ferguson04}
{Ferguson}, H.~C., {et~al.} 2004, \apjl, 600, L107

\bibitem[{{Finkelstein} {et~al.}(2014){Finkelstein}, {Ryan}, {Papovich},
  {Dickinson}, {Song}, {Somerville}, {Ferguson}, {Salmon}, {Giavalisco},
  {Koekemoer}, {Ashby}, {Behroozi}, {Castellano}, {Dunlop}, {Faber}, {Fazio},
  {Fontana}, {Grogin}, {Hathi}, {Jaacks}, {Kocevski}, {Livermore}, {McLure},
  {Merlin}, {Mobasher}, {Newman}, {Rafelski}, {Tilvi}, \&
  {Willner}}]{finkelstein14}
{Finkelstein}, S.~L., {et~al.} 2014, ArXiv e-prints

\bibitem[{{Finlator} {et~al.}(2012){Finlator}, {Oh}, {{\"O}zel}, \&
  {Dav{\'e}}}]{finlator12}
{Finlator}, K., {Oh}, S.~P., {{\"O}zel}, F., \& {Dav{\'e}}, R. 2012, \mnras,
  427, 2464

\bibitem[{{Giavalisco}(2002)}]{giavalisco02}
{Giavalisco}, M. 2002, \araa, 40, 579

\bibitem[{{Grazian} {et~al.}(2011){Grazian}, {Castellano}, {Koekemoer},
  {Fontana}, {Pentericci}, {Testa}, {Boutsia}, {Giallongo}, {Giavalisco}, \&
  {Santini}}]{grazian11}
{Grazian}, A., {et~al.} 2011, \aap, 532, A33

\bibitem[{{Grazian} {et~al.}(2012){Grazian}, {Castellano}, {Fontana},
  {Pentericci}, {Dunlop}, {McLure}, {Koekemoer}, {Dickinson}, {Faber},
  {Ferguson}, {Galametz}, {Giavalisco}, {Grogin}, {Hathi}, {Kocevski}, {Lai},
  {Newman}, \& {Vanzella}}]{grazian12}
---. 2012, \aap, 547, A51

\bibitem[{{Grillo} {et~al.}(2015){Grillo}, {Suyu}, {Rosati}, {Mercurio},
  {Balestra}, {Munari}, {Nonino}, {Caminha}, {Lombardi}, {De Lucia}, {Borgani},
  {Gobat}, {Biviano}, {Girardi}, {Umetsu}, {Coe}, {Koekemoer}, {Postman},
  {Zitrin}, {Halkola}, {Broadhurst}, {Sartoris}, {Presotto}, {Annunziatella},
  {Maier}, {Fritz}, {Vanzella}, \& {Frye}}]{grillo15}
{Grillo}, C., {et~al.} 2015, \apj, 800, 38

\bibitem[{{Hathi} {et~al.}(2008){Hathi}, {Jansen}, {Windhorst}, {Cohen},
  {Keel}, {Corbin}, \& {Ryan}}]{hathi08}
{Hathi}, N.~P., {Jansen}, R.~A., {Windhorst}, R.~A., {Cohen}, S.~H., {Keel},
  W.~C., {Corbin}, M.~R., \& {Ryan}, Jr., R.~E. 2008, \aj, 135, 156

\bibitem[{{Hayes} {et~al.}(2012){Hayes}, {Laporte}, {Pell{\'o}}, {Schaerer}, \&
  {Le Borgne}}]{hayes12}
{Hayes}, M., {Laporte}, N., {Pell{\'o}}, R., {Schaerer}, D., \& {Le Borgne},
  J.-F. 2012, \mnras, 425, L19

\bibitem[{{Huang} {et~al.}(2013){Huang}, {Ferguson}, {Ravindranath}, \&
  {Su}}]{huang13}
{Huang}, K.-H., {Ferguson}, H.~C., {Ravindranath}, S., \& {Su}, J. 2013, \apj,
  765, 68

\bibitem[{{Huang} {et~al.}(2015){Huang}, {Zheng}, {Wang}, {Ford}, {Lemze},
  {Moustakas}, {Shu}, {Van der Wel}, {Zitrin}, {Frye}, {Postman}, {Bartelmann},
  {Ben{\'{\i}}tez}, {Bradley}, {Broadhurst}, {Coe}, {Donahue}, {Infante},
  {Kelson}, {Koekemoer}, {Lahav}, {Medezinski}, {Moustakas}, {Rosati}, {Seitz},
  \& {Umetsu}}]{huang15}
{Huang}, X., {et~al.} 2015, \apj, 801, 12

\bibitem[{{Ishigaki} {et~al.}(2015{\natexlab{a}}){Ishigaki}, {Kawamata},
  {Ouchi}, {Oguri}, {Shimasaku}, \& {Ono}}]{ishigaki15}
{Ishigaki}, M., {Kawamata}, R., {Ouchi}, M., {Oguri}, M., {Shimasaku}, K., \&
  {Ono}, Y. 2015{\natexlab{a}}, \apj, 799, 12

\bibitem[{{Ishigaki} {et~al.}(2015{\natexlab{b}}){Ishigaki}, {Ouchi}, \&
  {Harikane}}]{ishigaki15b}
{Ishigaki}, M., {Ouchi}, M., \& {Harikane}, Y. 2015{\natexlab{b}}, ArXiv
  e-prints

\bibitem[{{Iwata} {et~al.}(2009){Iwata}, {Inoue}, {Matsuda}, {Furusawa},
  {Hayashino}, {Kousai}, {Akiyama}, {Yamada}, {Burgarella}, \&
  {Deharveng}}]{iwata09}
{Iwata}, I., {et~al.} 2009, \apj, 692, 1287

\bibitem[{{Jaacks} {et~al.}(2012){Jaacks}, {Choi}, {Nagamine}, {Thompson}, \&
  {Varghese}}]{jaacks12}
{Jaacks}, J., {Choi}, J.-H., {Nagamine}, K., {Thompson}, R., \& {Varghese}, S.
  2012, \mnras, 420, 1606

\bibitem[{{Jauzac} {et~al.}(2014){Jauzac}, {Cl{\'e}ment}, {Limousin},
  {Richard}, {Jullo}, {Ebeling}, {Atek}, {Kneib}, {Knowles}, {Natarajan},
  {Eckert}, {Egami}, {Massey}, \& {Rexroth}}]{jauzac14b}
{Jauzac}, M., {et~al.} 2014, \mnras, 443, 1549

\bibitem[{{Jauzac} {et~al.}(2015){Jauzac}, {Richard}, {Jullo}, {Cl{\'e}ment},
  {Limousin}, {Kneib}, {Ebeling}, {Natarajan}, {Rodney}, {Atek}, {Massey},
  {Eckert}, {Egami}, \& {Rexroth}}]{jauzac15}
---. 2015, \mnras, 452, 1437

\bibitem[{{Johnson} {et~al.}(2014){Johnson}, {Sharon}, {Bayliss}, {Gladders},
  {Coe}, \& {Ebeling}}]{johnson14}
{Johnson}, T.~L., {Sharon}, K., {Bayliss}, M.~B., {Gladders}, M.~D., {Coe}, D.,
  \& {Ebeling}, H. 2014, ArXiv e-prints

\bibitem[{{Jullo} \& {Kneib}(2009)}]{jullo09}
{Jullo}, E., \& {Kneib}, J.-P. 2009, \mnras, 395, 1319

\bibitem[{{Jullo} {et~al.}(2007){Jullo}, {Kneib}, {Limousin},
  {El{\'{\i}}asd{\'o}ttir}, {Marshall}, \& {Verdugo}}]{jullo07}
{Jullo}, E., {Kneib}, J.-P., {Limousin}, M., {El{\'{\i}}asd{\'o}ttir}, {\'A}.,
  {Marshall}, P.~J., \& {Verdugo}, T. 2007, New Journal of Physics, 9, 447

\bibitem[{{Karman} {et~al.}(2015){Karman}, {Caputi}, {Grillo}, {Balestra},
  {Rosati}, {Vanzella}, {Coe}, {Christensen}, {Koekemoer}, {Kr{\"u}hler},
  {Lombardi}, {Mercurio}, {Nonino}, \& {van der Wel}}]{karman15}
{Karman}, W., {et~al.} 2015, \aap, 574, A11

\bibitem[{{Kawamata} {et~al.}(2015){Kawamata}, {Ishigaki}, {Shimasaku},
  {Oguri}, \& {Ouchi}}]{kawamata15}
{Kawamata}, R., {Ishigaki}, M., {Shimasaku}, K., {Oguri}, M., \& {Ouchi}, M.
  2015, \apj, 804, 103

\bibitem[{{Kimm} \& {Cen}(2014)}]{kimm14}
{Kimm}, T., \& {Cen}, R. 2014, \apj, 788, 121

\bibitem[{Kinney {et~al.}(1996)Kinney, Calzetti, Bohlin, McQuade,
  Storchi-Bergmann, \& Schmitt}]{kinney96}
Kinney, A.~L., Calzetti, D., Bohlin, R.~C., McQuade, K., Storchi-Bergmann, T.,
  \& Schmitt, H.~R. 1996, The Astrophysical Journal, 467, 38

\bibitem[{{Kneib}(1993)}]{kneib93}
{Kneib}, J.-P. 1993, PhD thesis, Ph.~D.~thesis, Universit{\'e} Paul Sabatier,
  Toulouse, (1993)

\bibitem[{{Kneib} \& {Natarajan}(2011)}]{kneib11}
{Kneib}, J.-P., \& {Natarajan}, P. 2011, \aapr, 19, 47

\bibitem[{Krist {et~al.}(2011)Krist, Hook, \& Stoehr}]{Krist_Hook_Stoehr_2011}
Krist, J.~E., Hook, R.~N., \& Stoehr, F. 2011, {20 years of Hubble Space
  Telescope optical modeling using Tiny Tim} (SPIE - International Society for
  Optical Engineering), 81270J--81270J--16

\bibitem[{{Kuhlen} \& {Faucher-Gigu{\`e}re}(2012)}]{kuhlen12}
{Kuhlen}, M., \& {Faucher-Gigu{\`e}re}, C.-A. 2012, \mnras, 423, 862

\bibitem[{{Madau} {et~al.}(1999){Madau}, {Haardt}, \& {Rees}}]{madau99}
{Madau}, P., {Haardt}, F., \& {Rees}, M.~J. 1999, \apj, 514, 648

\bibitem[{{Mann} \& {Ebeling}(2012)}]{mann12}
{Mann}, A.~W., \& {Ebeling}, H. 2012, \mnras, 420, 2120

\bibitem[{{Mason} {et~al.}(2015){Mason}, {Trenti}, \& {Treu}}]{mason15}
{Mason}, C., {Trenti}, M., \& {Treu}, T. 2015, ArXiv e-prints

\bibitem[{{McLeod} {et~al.}(2015){McLeod}, {McLure}, {Dunlop}, {Robertson},
  {Ellis}, \& {Targett}}]{mcleod15}
{McLeod}, D.~J., {McLure}, R.~J., {Dunlop}, J.~S., {Robertson}, B.~E., {Ellis},
  R.~S., \& {Targett}, T.~A. 2015, \mnras, 450, 3032

\bibitem[{{McLure} {et~al.}(2009){McLure}, {Cirasuolo}, {Dunlop}, {Foucaud}, \&
  {Almaini}}]{mclure09}
{McLure}, R.~J., {Cirasuolo}, M., {Dunlop}, J.~S., {Foucaud}, S., \& {Almaini},
  O. 2009, \mnras, 395, 2196

\bibitem[{{McLure} {et~al.}(2013){McLure}, {Dunlop}, {Bowler}, {Curtis-Lake},
  {Schenker}, {Ellis}, {Robertson}, {Koekemoer}, {Rogers}, {Ono}, {Ouchi},
  {Charlot}, {Wild}, {Stark}, {Furlanetto}, {Cirasuolo}, \&
  {Targett}}]{mclure13}
{McLure}, R.~J., {et~al.} 2013, \mnras, 432, 2696

\bibitem[{{Medezinski} {et~al.}(2015){Medezinski}, {Umetsu}, {Okabe}, {Nonino},
  {Molnar}, {Massey}, {Dupke}, \& {Merten}}]{medezinski15}
{Medezinski}, E., {Umetsu}, K., {Okabe}, N., {Nonino}, M., {Molnar}, S.,
  {Massey}, R., {Dupke}, R., \& {Merten}, J. 2015, ArXiv e-prints

\bibitem[{{Merten} {et~al.}(2011){Merten}, {Coe}, {Dupke}, {Massey}, {Zitrin},
  {Cypriano}, {Okabe}, {Frye}, {Braglia}, {Jim{\'e}nez-Teja}, {Ben{\'{\i}}tez},
  {Broadhurst}, {Rhodes}, {Meneghetti}, {Moustakas}, {Sodr{\'e}}, {Krick}, \&
  {Bregman}}]{merten11}
{Merten}, J., {et~al.} 2011, \mnras, 417, 333

\bibitem[{{Mitchell-Wynne} {et~al.}(2015){Mitchell-Wynne}, {Cooray}, {Gong},
  {Ashby}, {Dolch}, {Ferguson}, {Finkelstein}, {Grogin}, {Kocevski},
  {Koekemoer}, {Primack}, \& {Smidt}}]{mitchell15}
{Mitchell-Wynne}, K., {et~al.} 2015, ArXiv e-prints

\bibitem[{{Montes} \& {Trujillo}(2014)}]{montes14}
{Montes}, M., \& {Trujillo}, I. 2014, \apj, 794, 137

\bibitem[{{Mosleh} {et~al.}(2012){Mosleh}, {Williams}, {Franx}, {Gonzalez},
  {Bouwens}, {Oesch}, {Labbe}, {Illingworth}, \& {Trenti}}]{mosleh12}
{Mosleh}, M., {et~al.} 2012, \apjl, 756, L12

\bibitem[{{Nestor} {et~al.}(2013){Nestor}, {Shapley}, {Kornei}, {Steidel}, \&
  {Siana}}]{nestor13}
{Nestor}, D.~B., {Shapley}, A.~E., {Kornei}, K.~A., {Steidel}, C.~C., \&
  {Siana}, B. 2013, \apj, 765, 47

\bibitem[{{Oesch} {et~al.}(2010){Oesch}, {Bouwens}, {Illingworth}, {Carollo},
  {Franx}, {Labb{\'e}}, {Magee}, {Stiavelli}, {Trenti}, \& {van
  Dokkum}}]{oesch10a}
{Oesch}, P.~A., {et~al.} 2010, \apjl, 709, L16

\bibitem[{Oesch {et~al.}(2010)Oesch, Bouwens, Illingworth, Carollo, Franx,
  Labb{\'{e}}, Magee, Stiavelli, Trenti, \& van Dokkum}]{oesch10}
Oesch, P.~A., {et~al.} 2010, The Astrophysical Journal, 709, L16

\bibitem[{{Oesch} {et~al.}(2012){Oesch}, {Bouwens}, {Illingworth}, {Labb{\'e}},
  {Trenti}, {Gonzalez}, {Carollo}, {Franx}, {van Dokkum}, \& {Magee}}]{oesch12}
{Oesch}, P.~A., {et~al.} 2012, \apj, 745, 110

\bibitem[{{Oesch} {et~al.}(2014){Oesch}, {Bouwens}, {Illingworth}, {Labb{\'e}},
  {Smit}, {Franx}, {van Dokkum}, {Momcheva}, {Ashby}, {Fazio}, {Huang},
  {Willner}, {Gonzalez}, {Magee}, {Trenti}, {Brammer}, {Skelton}, \&
  {Spitler}}]{oesch14}
---. 2014, \apj, 786, 108

\bibitem[{{Ogrean} {et~al.}(2015){Ogrean}, {van Weeren}, {Jones}, {Clarke},
  {Sayers}, {Mroczkowski}, {Nulsen}, {Forman}, {Murray}, {Pandey-Pommier},
  {Randall}, {Churazov}, {Bonafede}, {Kraft}, {David}, {Andrade-Santos},
  {Merten}, {Zitrin}, {Umetsu}, {Goulding}, {Roediger}, {Bagchi}, {Bulbul},
  {Donahue}, {Ebeling}, {Johnston-Hollitt}, {Mason}, {Rosati}, \&
  {Vikhlinin}}]{ogrean15}
{Ogrean}, G., {et~al.} 2015, ArXiv e-prints

\bibitem[{{Ono} {et~al.}(2013){Ono}, {Ouchi}, {Curtis-Lake}, {Schenker},
  {Ellis}, {McLure}, {Dunlop}, {Robertson}, {Koekemoer}, {Bowler}, {Rogers},
  {Schneider}, {Charlot}, {Stark}, {Shimasaku}, {Furlanetto}, \&
  {Cirasuolo}}]{ono13}
{Ono}, Y., {et~al.} 2013, \apj, 777, 155

\bibitem[{{Ouchi} {et~al.}(2009){Ouchi}, {Mobasher}, {Shimasaku}, {Ferguson},
  {Fall}, {Ono}, {Kashikawa}, {Morokuma}, {Nakajima}, {Okamura}, {Dickinson},
  {Giavalisco}, \& {Ohta}}]{ouchi09}
{Ouchi}, M., {et~al.} 2009, \apj, 706, 1136

\bibitem[{{Owers} {et~al.}(2011){Owers}, {Randall}, {Nulsen}, {Couch}, {David},
  \& {Kempner}}]{owers11}
{Owers}, M.~S., {Randall}, S.~W., {Nulsen}, P.~E.~J., {Couch}, W.~J., {David},
  L.~P., \& {Kempner}, J.~C. 2011, \apj, 728, 27

\bibitem[{{Pawlik} {et~al.}(2009){Pawlik}, {Schaye}, \& {van
  Scherpenzeel}}]{pawlik09}
{Pawlik}, A.~H., {Schaye}, J., \& {van Scherpenzeel}, E. 2009, \mnras, 394,
  1812

\bibitem[{{P{\'e}nin} {et~al.}(2015){P{\'e}nin}, {Cuby}, {Cl{\'e}ment},
  {Hibon}, {Kneib}, {Cassata}, \& {Ilbert}}]{penin15}
{P{\'e}nin}, A., {Cuby}, J.-G., {Cl{\'e}ment}, B., {Hibon}, P., {Kneib}, J.-P.,
  {Cassata}, P., \& {Ilbert}, O. 2015, \aap, 577, A74

\bibitem[{{Pentericci} {et~al.}(2014){Pentericci}, {Vanzella}, {Fontana},
  {Castellano}, {Treu}, {Mesinger}, {Dijkstra}, {Grazian}, {Brada{\v c}},
  {Conselice}, {Cristiani}, {Dunlop}, {Galametz}, {Giavalisco}, {Giallongo},
  {Koekemoer}, {McLure}, {Maiolino}, {Paris}, \& {Santini}}]{pentericci14}
{Pentericci}, L., {et~al.} 2014, \apj, 793, 113

\bibitem[{{Planck Collaboration} {et~al.}(2015){Planck Collaboration}, {Ade},
  {Aghanim}, {Arnaud}, {Ashdown}, {Aumont}, {Baccigalupi}, {Banday},
  {Barreiro}, {Bartlett}, \& et~al.}]{planck15}
{Planck Collaboration} {et~al.} 2015, ArXiv e-prints

\bibitem[{{Postman} {et~al.}(2012){Postman}, {Coe}, {Ben{\'{\i}}tez},
  {Bradley}, {Broadhurst}, {Donahue}, {Ford}, {Graur}, {Graves}, {Jouvel},
  {Koekemoer}, {Lemze}, {Medezinski}, {Molino}, {Moustakas}, {Ogaz}, {Riess},
  {Rodney}, {Rosati}, {Umetsu}, {Zheng}, {Zitrin}, {Bartelmann}, {Bouwens},
  {Czakon}, {Golwala}, {Host}, {Infante}, {Jha}, {Jimenez-Teja}, {Kelson},
  {Lahav}, {Lazkoz}, {Maoz}, {McCully}, {Melchior}, {Meneghetti}, {Merten},
  {Moustakas}, {Nonino}, {Patel}, {Reg{\"o}s}, {Sayers}, {Seitz}, \& {Van der
  Wel}}]{postman12}
{Postman}, M., {et~al.} 2012, \apjs, 199, 25

\bibitem[{{Rawle} {et~al.}(2015){Rawle}, {Altieri}, {Egami},
  {P{\'e}rez-Gonz{\'a}lez}, {Boone}, {Clement}, {Ivison}, {Richard},
  {Rujopakarn}, {Valtchanov}, {Walth}, {Weiner}, {Blain}, {Dessauges-Zavadsky},
  {Kneib}, {Lutz}, {Rodighiero}, {Schaerer}, \& {Smail}}]{rawle15}
{Rawle}, T.~D., {et~al.} 2015, ArXiv e-prints

\bibitem[{{Reddy} \& {Steidel}(2009)}]{reddy09}
{Reddy}, N.~A., \& {Steidel}, C.~C. 2009, \apj, 692, 778

\bibitem[{{Richard} {et~al.}(2014){Richard}, {Jauzac}, {Limousin}, {Jullo},
  {Cl{\'e}ment}, {Ebeling}, {Kneib}, {Atek}, {Natarajan}, {Egami}, {Livermore},
  \& {Bower}}]{richard14}
{Richard}, J., {et~al.} 2014, ArXiv e-prints

\bibitem[{{Richard} {et~al.}(2015){Richard}, {Patricio}, {Martinez}, {Bacon},
  {Cl{\'e}ment}, {Weilbacher}, {Soto}, {Wisotzki}, {Vernet}, {Pello}, {Schaye},
  {Turner}, \& {Martinsson}}]{richard14b}
---. 2015, \mnras, 446, L16

\bibitem[{{Robertson} {et~al.}(2014){Robertson}, {Ellis}, {Dunlop}, {McLure},
  {Stark}, \& {McLeod}}]{robertson14}
{Robertson}, B.~E., {Ellis}, R.~S., {Dunlop}, J.~S., {McLure}, R.~J., {Stark},
  D.~P., \& {McLeod}, D. 2014, ArXiv e-prints

\bibitem[{{Robertson} {et~al.}(2015){Robertson}, {Ellis}, {Furlanetto}, \&
  {Dunlop}}]{robertson15}
{Robertson}, B.~E., {Ellis}, R.~S., {Furlanetto}, S.~R., \& {Dunlop}, J.~S.
  2015, \apjl, 802, L19

\bibitem[{{Rodney} {et~al.}(2015){Rodney}, {Patel}, {Scolnic}, {Foley},
  {Molino}, {Brammer}, {Jauzac}, {Bradac}, {Coe}, {Broadhurst}, {Diego},
  {Graur}, {Hjorth}, {Hoag}, {Jha}, {Johnson}, {Kelly}, {Lam}, {McCully},
  {Medezinski}, {Meneghetti}, {Merten}, {Richard}, {Riess}, {Sharon},
  {Strolger}, {Treu}, {Wang}, {Williams}, \& {Zitrin}}]{rodney15}
{Rodney}, S.~A., {et~al.} 2015, ArXiv e-prints

\bibitem[{{Schaerer} \& {de Barros}(2009)}]{schaerer09}
{Schaerer}, D., \& {de Barros}, S. 2009, \aap, 502, 423

\bibitem[{{Schechter}(1976)}]{schechter76}
{Schechter}, P. 1976, \apj, 203, 297

\bibitem[{{Schenker} {et~al.}(2013{\natexlab{a}}){Schenker}, {Ellis},
  {Konidaris}, \& {Stark}}]{schenker13}
{Schenker}, M.~A., {Ellis}, R.~S., {Konidaris}, N.~P., \& {Stark}, D.~P.
  2013{\natexlab{a}}, \apj, 777, 67

\bibitem[{{Schenker} {et~al.}(2014{\natexlab{a}}){Schenker}, {Ellis},
  {Konidaris}, \& {Stark}}]{schenker14}
---. 2014{\natexlab{a}}, \apj, 795, 20

\bibitem[{{Schenker} {et~al.}(2014{\natexlab{b}}){Schenker}, {Ellis},
  {Konidaris}, \& {Stark}}]{schenker14b}
---. 2014{\natexlab{b}}, \apj, 795, 20

\bibitem[{{Schenker} {et~al.}(2013{\natexlab{b}}){Schenker}, {Robertson},
  {Ellis}, {Ono}, {McLure}, {Dunlop}, {Koekemoer}, {Bowler}, {Ouchi},
  {Curtis-Lake}, {Rogers}, {Schneider}, {Charlot}, {Stark}, {Furlanetto}, \&
  {Cirasuolo}}]{schenker13b}
{Schenker}, M.~A., {et~al.} 2013{\natexlab{b}}, \apj, 768, 196

\bibitem[{{Schiminovich} {et~al.}(2005){Schiminovich}, {Ilbert}, {Arnouts},
  {Milliard}, {Tresse}, {Le F{\`e}vre}, {Treyer}, {Wyder}, {Budav{\'a}ri},
  {Zucca}, {Zamorani}, {Martin}, {Adami}, {Arnaboldi}, {Bardelli}, {Barlow},
  {Bianchi}, {Bolzonella}, {Bottini}, {Byun}, {Cappi}, {Contini}, {Charlot},
  {Donas}, {Forster}, {Foucaud}, {Franzetti}, {Friedman}, {Garilli},
  {Gavignaud}, {Guzzo}, {Heckman}, {Hoopes}, {Iovino}, {Jelinsky}, {Le Brun},
  {Lee}, {Maccagni}, {Madore}, {Malina}, {Marano}, {Marinoni}, {McCracken},
  {Mazure}, {Meneux}, {Morrissey}, {Neff}, {Paltani}, {Pell{\`o}}, {Picat},
  {Pollo}, {Pozzetti}, {Radovich}, {Rich}, {Scaramella}, {Scodeggio},
  {Seibert}, {Siegmund}, {Small}, {Szalay}, {Vettolani}, {Welsh}, {Xu}, \&
  {Zanichelli}}]{schiminovich05}
{Schiminovich}, D., {et~al.} 2005, \apjl, 619, L47

\bibitem[{{Schirmer} {et~al.}(2015){Schirmer}, {Carrasco}, {Pessev}, {Garrel},
  {Winge}, {Neichel}, \& {Vidal}}]{schirmer15}
{Schirmer}, M., {Carrasco}, E.~R., {Pessev}, P., {Garrel}, V., {Winge}, C.,
  {Neichel}, B., \& {Vidal}, F. 2015, \apjs, 217, 33

\bibitem[{{Schmidt} {et~al.}(2014){Schmidt}, {Treu}, {Trenti}, {Bradley},
  {Kelly}, {Oesch}, {Holwerda}, {Shull}, \& {Stiavelli}}]{schmidt14}
{Schmidt}, K.~B., {et~al.} 2014, \apj, 786, 57

\bibitem[{{Shapley} {et~al.}(2006){Shapley}, {Steidel}, {Pettini},
  {Adelberger}, \& {Erb}}]{shapley06}
{Shapley}, A.~E., {Steidel}, C.~C., {Pettini}, M., {Adelberger}, K.~L., \&
  {Erb}, D.~K. 2006, \apj, 651, 688

\bibitem[{{Siana} {et~al.}(2010){Siana}, {Teplitz}, {Ferguson}, {Brown},
  {Giavalisco}, {Dickinson}, {Chary}, {de Mello}, {Conselice}, {Bridge},
  {Gardner}, {Colbert}, \& {Scarlata}}]{siana10}
{Siana}, B., {et~al.} 2010, \apj, 723, 241

\bibitem[{{Soucail} {et~al.}(1987){Soucail}, {Fort}, {Mellier}, \&
  {Picat}}]{soucail87}
{Soucail}, G., {Fort}, B., {Mellier}, Y., \& {Picat}, J.~P. 1987, \aap, 172,
  L14

\bibitem[{{Stark} {et~al.}(2010){Stark}, {Ellis}, {Chiu}, {Ouchi}, \&
  {Bunker}}]{stark10}
{Stark}, D.~P., {Ellis}, R.~S., {Chiu}, K., {Ouchi}, M., \& {Bunker}, A. 2010,
  \mnras, 408, 1628

\bibitem[{Steidel {et~al.}(1996)Steidel, Giavalisco, Pettini, Dickinson, \&
  Adelberger}]{steidel96}
Steidel, C.~C., Giavalisco, M., Pettini, M., Dickinson, M., \& Adelberger,
  K.~L. 1996, The Astrophysical Journal, 462, L17

\bibitem[{{Steidel} {et~al.}(2001){Steidel}, {Pettini}, \&
  {Adelberger}}]{steidel01}
{Steidel}, C.~C., {Pettini}, M., \& {Adelberger}, K.~L. 2001, \apj, 546, 665

\bibitem[{{Trenti} {et~al.}(2011){Trenti}, {Bradley}, {Stiavelli}, {Oesch},
  {Treu}, {Bouwens}, {Shull}, {MacKenty}, {Carollo}, \&
  {Illingworth}}]{trenti11}
{Trenti}, M., {et~al.} 2011, \apjl, 727, L39

\bibitem[{{Treu} {et~al.}(2013){Treu}, {Schmidt}, {Trenti}, {Bradley}, \&
  {Stiavelli}}]{treu13}
{Treu}, T., {Schmidt}, K.~B., {Trenti}, M., {Bradley}, L.~D., \& {Stiavelli},
  M. 2013, \apjl, 775, L29

\bibitem[{{Treu} {et~al.}(2015){Treu}, {Schmidt}, {Brammer}, {Vulcani}, {Wang},
  {Brada{\v c}}, {Dijkstra}, {Dressler}, {Fontana}, {Gavazzi}, {Henry}, {Hoag},
  {Huang}, {Jones}, {Kelly}, {Malkan}, {Mason}, {Pentericci}, {Poggianti},
  {Stiavelli}, {Trenti}, \& {von der Linden}}]{treu15}
{Treu}, T., {et~al.} 2015, ArXiv e-prints

\bibitem[{{van der Wel} {et~al.}(2011){van der Wel}, {Straughn}, {Rix},
  {Finkelstein}, {Koekemoer}, {Weiner}, {Wuyts}, {Bell}, {Faber}, {Trump},
  {Koo}, {Ferguson}, {Scarlata}, {Hathi}, {Dunlop}, {Newman}, {Dickinson},
  {Jahnke}, {Salmon}, {de Mello}, {Kocevski}, {Lai}, {Grogin}, {Rodney}, {Guo},
  {McGrath}, {Lee}, {Barro}, {Huang}, {Riess}, {Ashby}, \&
  {Willner}}]{vanderwel11}
{van der Wel}, A., {et~al.} 2011, \apj, 742, 111

\bibitem[{{Vanzella} {et~al.}(2009){Vanzella}, {Giavalisco}, {Dickinson},
  {Cristiani}, {Nonino}, {Kuntschner}, {Popesso}, {Rosati}, {Renzini}, {Stern},
  {Cesarsky}, {Ferguson}, \& {Fosbury}}]{vanzella09}
{Vanzella}, E., {et~al.} 2009, \apj, 695, 1163

\bibitem[{{Wang} {et~al.}(2015){Wang}, {Hoag}, {Huang}, {Treu}, {Bradac},
  {Schmidt}, {Brammer}, {Vulcani}, {Jones}, {Ryan}, {Amorin}, {Castellano},
  {Fontana}, {Merlin}, \& {Trenti}}]{wang15}
{Wang}, X., {et~al.} 2015, ArXiv e-prints

\bibitem[{{Wilkins} {et~al.}(2010){Wilkins}, {Bunker}, {Ellis}, {Stark},
  {Stanway}, {Chiu}, {Lorenzoni}, \& {Jarvis}}]{wilkins10}
{Wilkins}, S.~M., {Bunker}, A.~J., {Ellis}, R.~S., {Stark}, D., {Stanway},
  E.~R., {Chiu}, K., {Lorenzoni}, S., \& {Jarvis}, M.~J. 2010, \mnras, 403, 938

\bibitem[{{Wilkins} {et~al.}(2013){Wilkins}, {Coulton}, {Caruana}, {Croft},
  {Matteo}, {Khandai}, {Feng}, {Bunker}, \& {Elbert}}]{wilkins13}
{Wilkins}, S.~M., {et~al.} 2013, \mnras, 435, 2885

\bibitem[{{Willott} {et~al.}(2013){Willott}, {McLure}, {Hibon}, {Bielby},
  {McCracken}, {Kneib}, {Ilbert}, {Bonfield}, {Bruce}, \& {Jarvis}}]{willott13}
{Willott}, C.~J., {et~al.} 2013, \aj, 145, 4

\bibitem[{{Wise} {et~al.}(2014){Wise}, {Demchenko}, {Halicek}, {Norman},
  {Turk}, {Abel}, \& {Smith}}]{wise14}
{Wise}, J.~H., {Demchenko}, V.~G., {Halicek}, M.~T., {Norman}, M.~L., {Turk},
  M.~J., {Abel}, T., \& {Smith}, B.~D. 2014, \mnras, 442, 2560

\bibitem[{{Wong} {et~al.}(2012){Wong}, {Ammons}, {Keeton}, \&
  {Zabludoff}}]{wong12}
{Wong}, K.~C., {Ammons}, S.~M., {Keeton}, C.~R., \& {Zabludoff}, A.~I. 2012,
  \apj, 752, 104

\bibitem[{{Zheng} {et~al.}(2014){Zheng}, {Shu}, {.~Moustakas}, {Zitrin},
  {Ford}, {Huang}, {Broadhurst}, {Molino}, {Diego}, {Infante}, {Bauer}, \&
  {Kelson}}]{zheng14}
{Zheng}, W., {et~al.} 2014, ArXiv e-prints

\bibitem[{{Zitrin} {et~al.}(2011){Zitrin}, {Broadhurst}, {Coe}, {Liesenborgs},
  {Ben{\'{\i}}tez}, {Rephaeli}, {Ford}, \& {Umetsu}}]{zitrin11}
{Zitrin}, A., {Broadhurst}, T., {Coe}, D., {Liesenborgs}, J., {Ben{\'{\i}}tez},
  N., {Rephaeli}, Y., {Ford}, H., \& {Umetsu}, K. 2011, \mnras, 413, 1753

\bibitem[{{Zitrin} {et~al.}(2015){Zitrin}, {Ellis}, {Belli}, \&
  {Stark}}]{zitrin15}
{Zitrin}, A., {Ellis}, R.~S., {Belli}, S., \& {Stark}, D.~P. 2015, \apjl, 805,
  L7

\bibitem[{{Zitrin} {et~al.}(2014){Zitrin}, {Zheng}, {Broadhurst}, {Moustakas},
  {Lam}, {Shu}, {Huang}, {Diego}, {Ford}, {Lim}, {Bauer}, {Infante}, {Kelson},
  \& {Molino}}]{zitrin14}
{Zitrin}, A., {et~al.} 2014, ArXiv e-prints

\end{thebibliography}

\end{document}